\documentclass[12pt]{article}

\pdfoutput=1

\usepackage{placeins}
\usepackage{amsmath,amssymb,amsfonts,amsthm,bm}    
\usepackage{algorithm2e}
\usepackage{graphicx}
\usepackage{lineno}
\usepackage{pdfpages}
\usepackage[onehalfspacing]{setspace}
\usepackage[margin=1.5in]{geometry}
\usepackage{subcaption}
\usepackage{setspace}
\usepackage{placeins}
\usepackage{url}
\usepackage{xcolor}
\usepackage{mathtools}
\usepackage{titlesec}
\usepackage{natbib}
\usepackage{fixmath}
\usepackage{bbm}
\usepackage{booktabs}
\usepackage{multirow,bigdelim}
\usepackage{comment}
\usepackage{commath}

\titlelabel{\thetitle.\quad}

\begin{document}
	
\def\spacingset#1{\renewcommand{\baselinestretch}%
{#1}\small\normalsize} \spacingset{1}
	

\title{\bf Intensity Estimation on Geometric Networks with Penalized Splines}
\author{Marc Schneble and G\"oran Kauermann \thanks{We would like to thank the elite graduate program Data Science at LMU Munich and the Munich Center for Machine Learning (MCML) for funding.}
\hspace{.2cm}\\
Department of Statistics, Ludwig-Maximilians Universit\"at M\"unchen}
		
\date{}	
		
\maketitle

\bigskip
\begin{abstract}
\noindent 
In the past decades, the growing amount of network data has lead to many novel statistical models. In this paper we consider so called geometric networks. Typical examples are road networks or other infrastructure networks. But also the neurons or the blood vessels in a human body can be interpreted as a geometric network embedded in a three-dimensional space. In all these applications a network specific metric rather than the Euclidean metric is usually used, which makes the analyses on network data challenging. We consider network based point processes and our task is to estimate the intensity (or density) of the process which allows to detect high- and low- intensity regions of the underlying stochastic processes. Available routines that tackle this problem are commonly based on kernel smoothing methods. However, kernel based estimation in general exhibits some drawbacks such as suffering from boundary effects and the locality of the smoother. In an Euclidean space, the disadvantages of kernel methods can be overcome by using penalized spline smoothing. We here extend penalized spline smoothing towards smooth intensity estimation on geometric networks and apply the approach to both, simulated and real world data. The results show that penalized spline based intensity estimation is numerically efficient and outperforms kernel based methods. Furthermore, our approach easily allows to incorporate covariates, which allows to respect the network geometry in a regression model framework. 
\end{abstract}
	
\noindent%

\section{Introduction}
In statistical network analysis, a (static) network is usually considered as a graph that is characterized by a set of vertices which are connected by a set of, possibly weighted, edges. In this matter, the interest usually lies in the mutual relationship and the dependencies of the vertices, sometimes called ``actors'' \citep{snijders1996stochastic}, that are induced by the edges. For a general overview, see e.g. \cite{goldenberg2010survey}. In the context of this paper, a network is rather considered as a geometric object embedded in an Euclidean space. We use the term ``geometric network'' and a typical example is a network of streets. The setting is that we observe a spatial point process on the network edges and focus is on estimating the intensity (or density) of this process.

Regarding the data structure, the question arises why one should analyze data points on a geometric network and not in the Euclidean space itself. To illustrate this, consider a point pattern that seems to be clustered in the plane. However, the points might just be uniformly distributed on a network where many network segments are clustered within a small area. A typical example is the distribution of traffic accidents in an urban area \citep{mcswiggan2017kernel}. Theoretically, such events can only occur on a network of streets which is often considered being embedded in the plane. Therefore, the statistical analyses of point patterns which are distributed across an Euclidean space and point patterns which are distributed only on a geometric network are tremendously different. 

Due to the increasing availability of network based data, the last 25 years have seen a broad range of literature that is concerned with network based point processes. Amongst the first statistical analyses of spatial point patterns on a network were proposed by \cite{okabe1995statistical}, \cite{okabe2001k} and \cite{spooner2004spatial}. They all noted that in the context of geometric network data the Euclidean distance needs to be replaced by the shortest path distance in order to respect the network geometry. This resulted e.g. in the geometrically corrected network K-function, a modified version of Ripley's K-function in two dimensions. The network K-function can be used to analyze the correlation structure of point patterns on a network \citep{ang2012geometrically}.  \cite{baddeley2015spatial} discuss the topic in general, and from an application point of view, including the R package \texttt{spatstat}, which is also maintained by the same authors. Among other contributions, a huge library of functions is provided in order to create, manipulate and analyze both, a point pattern on a network, which is embedded in the plane, and the network itself. Furthermore, \cite{baddeley2015spatial} also treat marked point processes, intensities of point processes depending on covariates and point processes on trees which are networks without loops. Marked point processes on directed linear networks are further discussed by \cite{rasmussen2018point} for various kinds of stochastic point processes.

\cite{borruso2008network} and \cite{xie2008kernel} developed kernel density estimation on a network geometry which is performed by exploiting the shortest path distance. However, both articles did not take into account that around vertices with more than two adjoining segments, there is more network mass within a certain shortest path distance. Hence, this approach leads to biased estimates, especially if the point pattern is distributed according to a uniform distribution on the network. \cite{okabe2009kernel} solved this problem by introducing equal-split (dis-)continuous  kernel density estimation. The idea is to split the mass of the kernel functions equally across all other segments that depart from a vertex when approaching this vertex from one side. The estimation procedure, including automatic bandwidth selection, is implemented in the R package \texttt{spatstat}. The approach was refined in \cite{mcswiggan2017kernel}, also including automatic bandwidth selection. Instead of a finite sum of paths over the network, they consider an infinite sum leading to a diffusion estimate that can be computed via a heat equation on the network. Furthermore, \cite{moradi2018kernel} showed in their application that an extension of Diggle's \citep{diggle1985kernel} non-parametric edge-corrected kernel-based intensity estimator is superior to the equal-split discontinuous estimator that was proposed by \cite{okabe2009kernel}. Recently, fused density estimation was proposed by \cite{bassett2018fused} to estimate the density on a geometric network. This estimator is the solution of a total variation regularized maximum-likelihood  problem. However, the method does not produce continuous density estimates. In the special case of a river network, which results as a directed and acyclic network, \cite{o2014flexible} used penalized piecewise constant functions to estimate the river flow, which can be interpreted as penalized splines (P-splines,  \citeauthor{eilers1996flexible}, \citeyear{eilers1996flexible}) of order 0. Hence, the density estimate here is also not continuous. The paper was reviewed by \cite{rushworth2015validation} who implemented the theory in the R package \texttt{smnet}.  

Amongst others, \cite{eilers1996flexible} argued that the kernel density estimators of points on the real line suffer from boundary effects, e.g. if the domain of the data is not specified correctly. Instead, the authors proposed to estimate the density by making use of penalized splines in order to smooth the histogram that is created by binning the data with small bin widths. In the course of time, this concept has been extended, among others, to allow for density estimation of multiple dimensional data \citep{currie2006generalized} or to represent the density as a mixture of weighted penalized spline densities \citep{schellhase2012density}. A comprehensive survey of penalized spline theory and its application is given in \cite{eilers2015twenty}. Generally, penalized spline estimation has become a major workhorse in statistical modeling as demonstrated in \citeauthor{ruppert2003semiparametric} (\citeyear{ruppert2003semiparametric}, \citeyear{ruppert2009semiparametric}).

In this paper we extend the penalized spline approach to work on geometric networks. The focus is to estimate the intensity (or density) of a point process on the network given realizations of this process as data. The remainder of this paper is structured as follows. Section \ref{sec: notation}  introduces some basic notation related to network graphs \citep{kolaczyk2014statistical}, geometric networks and stochastic point processes on networks. Section \ref{sec: methodology} treats our new methodology to estimate the intensity of a point process on a geometric network with penalized splines which is followed by some examples exploiting both, simulated data (Section \ref{sec: simulation}) and real world data (Section \ref{sec: application}). In Section \ref{sec: discussion} we discuss the results of this paper and present some prospects of how our approach can be extended and generalized.

\section{Notation and Problem}
\label{sec: notation}

Consider a set $V = \lbrace v_1,\dots,v_W\rbrace$ of $W\in\mathbb{N}$ elements which we call vertices. Further, let $E = \lbrace e_1,\dots,e_M \rbrace \subset V \times V$ be a set of $M \in \mathbb{N}$ pairs $e_m = (v_{i}, v_{j})$ which we call edges. Putting these together leads to the network graph $L = (V, E)$ and we denote $L$ as the graph representation of the network defined by a set of vertices $V$ and a set of edges $E$. In this paper, we only consider undirected networks, i.e. there is an edge from $v_i$ to $v_j$ if and only if there is an edge from $v_j$ to $v_i$. An edge $e$ is called incident to a vertex $v$ if there is another vertex $v_i  \in V (v_i \neq v)$ such that $e = (v, v_i) \in E$ . The degree of a vertex $v$, denoted by $\text{deg}(v)$ is defined as the count of edges which are incident to $v$ and for our purpose we always remove a vertex $v$ from $V$ if $\text{deg}(v) = 0$. 

A geometric network also typically exhibits a geometric representation as a subset of an Euclidean space $\mathbb{R}^q$ for $q \geq 2$. In this case, the set of vertices $V$ in the network graph representation can be viewed as a set $\mathbold{V} = \lbrace \mathbold{v}_1,\dots,\mathbold{v}_W \rbrace$ of vectors with $\mathbold{v}_i \in \mathbb{R}^q$ for $i = 1,\dots,W$. Consequently, the edges can be viewed as a set $\mathbold{E} = \lbrace \mathbold{e_1},\dots,\mathbold{e}_M \rbrace$ of line or curve segments with each $\mathbold{e}_m \subset \mathbb{R}^q$ being the connection between two vertices $\mathbold{v}_{i}$ and $\mathbold{v}_{j}$. If $\mathbold{e}_m$ is a straight line in $\mathbb{R}^q$ between its endpoints $\mathbold{v}_i$ and $\mathbold{v}_j$, then $\mathbold{e}_m = \lbrace  \mathbold{v}_i + (1-t)\mathbold{v_j} \mid 0 \leq t \leq 1 \rbrace$ and the length $d_m$ of the line $\mathbold{e}_m$ is equal to the Euclidean distance between $\mathbold{v}_i$ and $\mathbold{v}_j$ in $\mathbb{R}^q$, i.e. $d_m = |\mathbold{e}_m| = ||\mathbold{v}_i - \mathbold{v}_j||_2$. More generally, such an edge can be rather described by the image set $\mathbold{e}_m =  \nu_m([a_m, b_m])$ of a parametric curve \citep{heuser2006lehrbuch} $\nu_m : [a_m,b_m] \rightarrow \mathbb{R}^q$ with $a_m<b_m, \nu_m(a_m) = \mathbold{v}_i$ and $\nu_m(b_m) = \mathbold{v}_j$. Then, the length of the curve segment $\mathbold{e}_m$ is given by
\begin{equation}
d_m = |\mathbold{e}_m| = \lim_{N \rightarrow \infty} \sum_{i=1}^N ||\nu_m(t_i) - \nu_m(t_{i-1})||_2 = \int_a^b ||\nu_m'(t)||_2 dt,
\label{eq: arc length}
\end{equation}
with $t_i = a_m+i(b_m-a_m)/N$ for $i = 1, \dots,N$. The last equation in \eqref{eq: arc length} holds if the parametric curve $\nu_m$ is (piecewise) differentiable on $[a_m,b_m]$, which we generally assume throughout the paper. The distance measure defined in \eqref{eq: arc length} is usually denoted as arc length of the curve $\nu_m(\cdot)$ and is independent from its parameterization. 

We can now define the geometric representation of a network graph $L$ as $\mathbold{L} = \bigcup_{m=1}^M \mathbold{e}_m \subset \mathbb{R}^q$. Hence, there is a one-to-one correspondence between $L$ and $\mathbold{L}$ which we exploit consistently in this paper. According to the network graph representation, we also define a vertex degree for the geometric network representation, meaning that $\text{deg}(\mathbold{v})$ denotes the count of segments which have an endpoint equal to $\mathbold{v}$. Furthermore, the lengths $d_m$ of the curves $\mathbold{e}_m$ from  \eqref{eq: arc length} imply a metric $d_\mathbold{L}: \mathbold{L} \times \mathbold{L} \rightarrow [0, \infty)$ on $\mathbold{L}$. More precisely, $d_\mathbold{L}(\mathbold{z}_1, \mathbold{z}_2)$ denotes the shortest path distance between two points $\mathbold{z}_1, \mathbold{z}_2$ on $\mathbold{L}$ and with $[\mathbold{z}_1; \mathbold{z}_2] \subset \mathbold{L}$ or with $[\mathbold{z}_1; \mathbold{z}_2) \subset \mathbold{L}$ we denote the corresponding path, where a round bracket indicates that an endpoint is not contained in the set. The total length of the geometric network is $|\mathbold{L}| = \sum_{m=1}^M d_m$. If the network is not connected, i.e. the corresponding network graph $L$ consists of more than one connected component, we can use the extended metric \citep{beer2013structure} $d_\mathbold{L} : \mathbold{L} \times \mathbold{L} \rightarrow [0, \infty) \cup \infty$. In this case, the same methodology can be applied unmodified.  

\begin{figure}
\center
\includegraphics[width=0.49\textwidth]{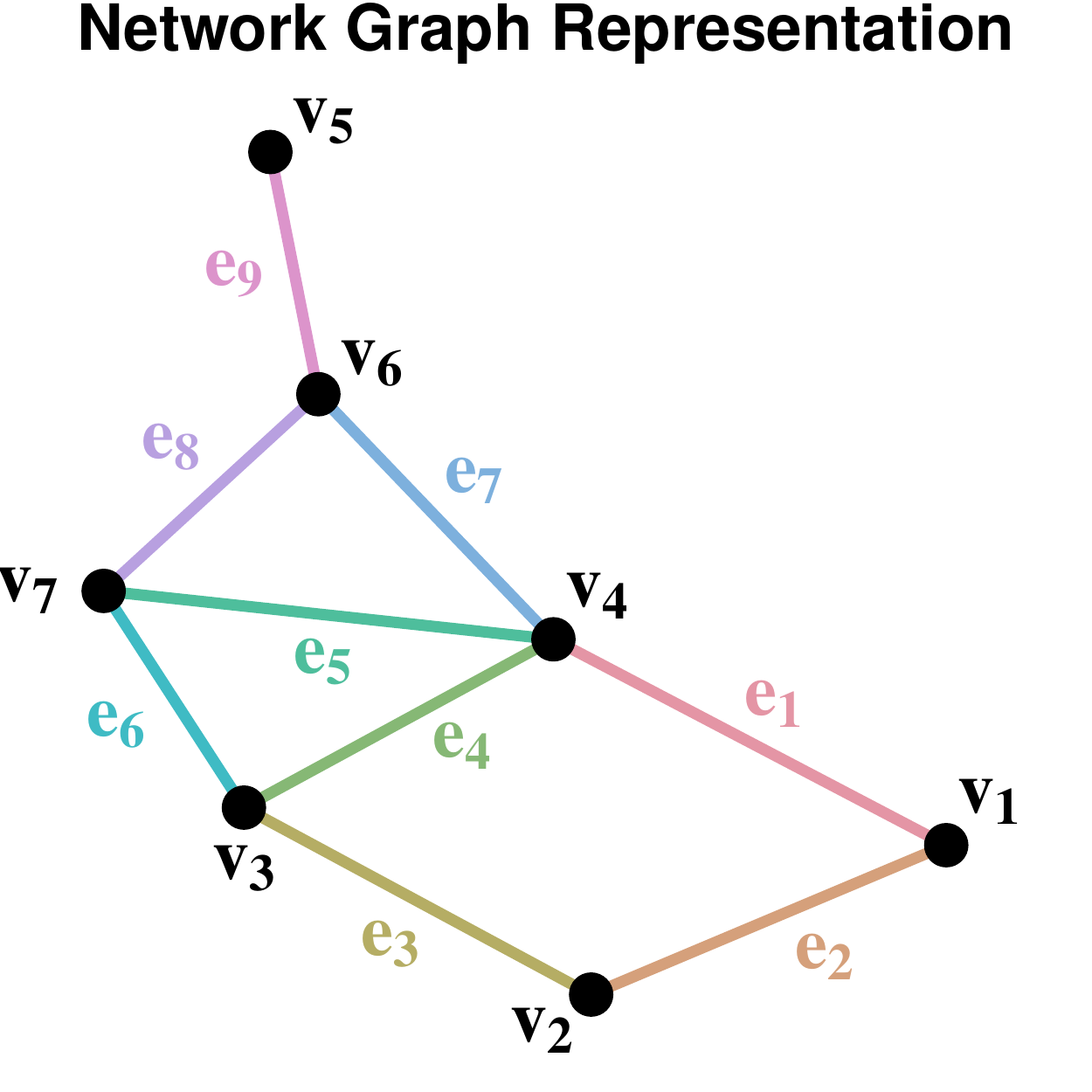} 
\includegraphics[width=0.49\textwidth]{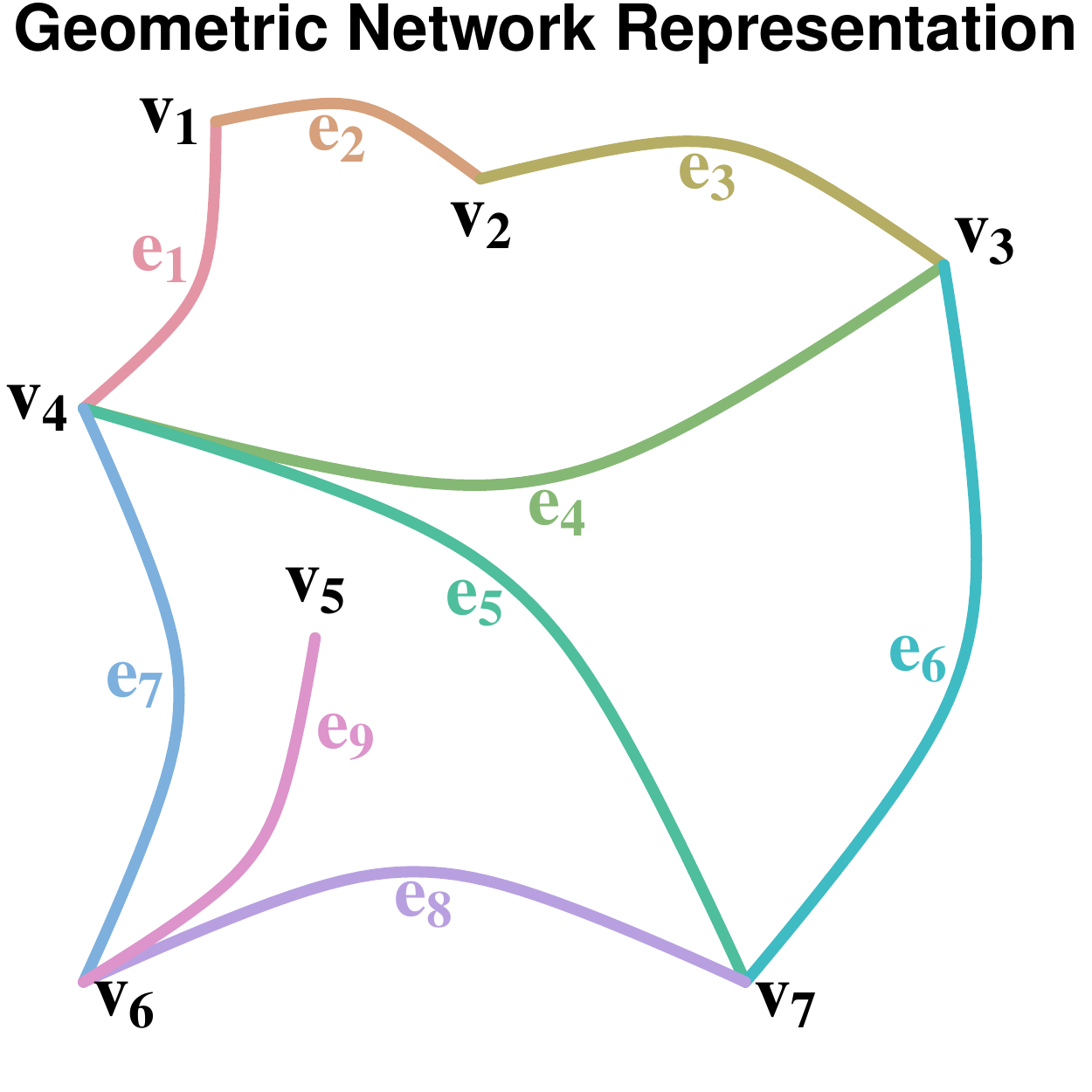} 
\caption{Two different representations of a network: Left panel: Network graph representation $L$; Right Panel: Geometric network representation $\mathbold{L}$.}
\label{fig: example notation}
\end{figure} 

Also note that the geometric representation $\mathbold{L}$ is not necessarily unique which can be seen from the following consideration: Let $v$ be a vertex (in network graph representation) with exactly two incident edges $e_m = (v_i, v)$ and $e_n = (v, v_j)$, i.e. $\text{deg}(v) = 2$ and $(v_i, v_j) \notin E$. If we remove $v$ from $V$ as well as $e_m, e_n$ from $E$ but add the edge $e = (v_i, v_j)$ to $E$, the network graph representation of $L = (V,E)$ has changed. In the geometric representation, we can remove $\mathbold{v}$ from $\mathbold{V}$ as well as $\mathbold{e}_m, \mathbold{e}_n$ from $\mathbold{E}$ and add the segment $\mathbold{e} = \mathbold{e}_m \cup \mathbold{e}_n$ to $\mathbold{E}$ which does not change $\mathbold{L}$. To exemplify the notation introduced in this section, Figure \ref{fig: example notation} shows a small network as a network graph (left panel) and as  a geometric network embedded in the plane (right panel). The figure also visualizes that the function of a vertex $\mathbold{v}$ in the geometric network representation is merely being the endpoint of $\text{deg}(\mathbold{v})$ curves. Hence, in this network the vertex $\mathbold{v}_2$ could be removed from $\mathbold{L}$ without changing its geometric representation. 

We now consider the following setting, see also \cite{mcswiggan2017kernel}. Let $\mathcal{X}$ be a stochastic point process on the geometric network $\mathbold{L}$ with continuous intensity $\varphi_\mathcal{X}: \mathbold{L} \rightarrow [0, \infty)$. The expected number of points in a set $\mathbold{K} \subset \mathbold{L}$ is then defined through
\begin{equation*}
\int_\mathbold{K} \varphi_\mathcal{X}(\mathbold{z})  \dif \mathbold{z} = \sum_{m=1}^M \int_{\mathbold{K} \subset \mathbold{e}_m} \varphi_\mathcal{X}(\mathbold{z}) \dif_{\text{ }|m}\mathbold{z},
\end{equation*}
where $\dif_{\text{ }|m}\mathbold{z}$ denotes integration with respect to the curve $\mathbold{e}_m$. Our aim is to estimate the intensity of the point process $\mathcal{X}$ on $\mathbold{L}$ given that we observe realizations $\mathbold{x}_1, \dots, \mathbold{x}_n$ of this process. The point process $\mathcal{X}$ can equivalently be defined through a density function $f_\mathcal{X} : \mathbold{L} \rightarrow [0,\infty)$. The probability that a random point $\mathbold{X}_i \sim f_\mathcal{X}$ falls into a subset $\mathbold{K} \subset \mathbold{L}$ is then given by $\mathbb{P}(\mathbold{X}_i \in \mathbold{K}) = \int_\mathbold{K} f_\mathcal{X}(\mathbold{z}) \dif \mathbold{z}$. 

\section{Methodology}
\label{sec: methodology}

\subsection{B-Splines on a Network}
First, we briefly review B-splines (compare \citeauthor{ruppert2003semiparametric}, \citeyear{ruppert2003semiparametric} or \citeauthor{fahrmeir2013regression}, \citeyear{fahrmeir2013regression}). To start, assume first a simple point process $\mathcal{X}$ with intensity $\varphi_\mathcal{X}(z)$,  where $z$ is univariate and takes values in the bounded interval  $[a,b]$. The goal is to estimate $\varphi_\mathcal{X}(z)$ in a smooth and flexible way. To do so, we approximate the logarithmized intensity $\nu_\mathcal{X} = \log \varphi_\mathcal{X}$  through a B-spline basis representation $\nu_\mathcal{X}(z) = \sum_{j=1}^J \gamma_j B_j^l(z)$, where $B_j^l(\cdot)$ are B-splines of order $l \in \mathbb{N}_0$ and $\mathbold{\gamma} = (\gamma_1,\dots,\gamma_J)^\top$ is a vector of regression coefficients that needs to be estimated from the data. For the construction of B-splines, we use $I$ interior knots $a = \tau_1 < \dots < \tau_I = b$. The $J = I+l-1$ basis functions $B_j^l(\cdot)$ are each locally supported on $l+2$ adjacent knots and can be calculated recursively from lower order basis functions \citep{de1972calculating}. An important property of a B-spline basis is that $\sum_{j=1}^J B_j^l(z) = 1$ holds for $z \in [a,b]$ and any order of B-splines $l$. This property also needs to be respected in the geometric network case.

Subsequently, we restrict ourselves to linear B-spline bases for simplicity of presentation. For simplicity of notation we drop the superscript $l$ in the B-spline notation,  i.e. we construct B-splines of order $l = 1$ on a geometric network $\mathbold{L}$. Such a basis can be constructed straightforwardly using the one-dimensional definitions from above. On every curve $\mathbold{e}_m$, which has endpoints $\mathbold{v}_i$ and $\mathbold{v}_j$, we specify an equidistant sequence of $I_m$ knots $\mathbold{v}_i =  \mathbold{\tau}_{m,1}, \dots, \mathbold{\tau}_{m,{I_m}} = \mathbold{v}_j$ with $\mathbold{\tau}_{m,k} \in \mathbold{e}_m$ for $k = 1,\dots,I_m$, where $d_\mathbold{L}(\mathbold{\tau}_{m,k}, \mathbold{\tau}_{m,{k-1}}) = \delta_m$. Note that a knot which is equal to a vertex $\mathbold{v}$ is contained in the knot sequence of $\text{deg}(\mathbold{v})$ segments but it is still the same knot. Unlike to the one-dimensional setup, it is in general not possible to choose the set of knots to be equidistant on the entire geometric network $\mathbold{L}$ with respect to all curve lengths $d_m$. However, we may choose a global knot distance $\delta$ such that it is close to an equidistant allocation of knots on the entire geometric network. Let therefore $\lceil \cdot \rceil$ denote the upwards rounded integer and $\lfloor \cdot \rfloor$ the corresponding downwards rounded integer. We then define
\begin{equation}
\delta_m = \begin{cases}
d_m / \lfloor \frac{d_m}{\delta} \rfloor, \quad \frac{d_m}{\delta} - \lfloor \frac{d_m}{\delta} \rfloor < 0.5 \\
d_m / \lceil \frac{d_m}{\delta} \rceil, \quad \frac{d_m}{\delta} - \lfloor \frac{d_m}{\delta} \rfloor \geq 0.5
\end{cases},
 \label{eq: delta_m}
\end{equation}  
which leads to curve-specific knot distances $\delta_m$ which are as similar as possible for a given overall knot distance $\delta$. Generally, we will choose $\delta$ rather small such that the differences between the $\delta_m$ are small and can be considered as negligible. This will become more clear later, when we also introduce a penalization component in the estimation. 

Having the set of knots defined as above, we can construct a linear B-spline basis $B$ on the geometric network $\mathbold{L}$. First, we use for every segment $\mathbold{e}_m$ with endpoints $\mathbold{v}_i$ and $\mathbold{v}_j$ the equidistant sequence of knots $\mathbold{v}_i = \mathbold{\tau}_{m,1},\dots,\mathbold{\tau}_{m,{I_m}} = \mathbold{v}_j$ from above to construct $J_m = I_m-2$ linear B-splines 
$B_{m,1},\dots,B_{m,{J_m}}$. These B-splines are defined accordingly to the univariate case by 
\begin{equation}
B_{m,k}(\mathbold{z}) = \frac{d_\mathbold{L}(\mathbold{z}, \mathbold{\tau}_{m,k})}{\delta_m} \mathbbm{1}_{[\mathbold{\tau}_{m,k}, \mathbold{\tau}_{m,k+1})}(\mathbold{z}) + \frac{d_\mathbold{L}(\mathbold{\tau}_{m,k+2}, \mathbold{z})}{\delta_m} \mathbbm{1}_{[\mathbold{\tau}_{m,k+1}, \mathbold{\tau}_{m,k+2})}(\mathbold{z}) 
\label{eq: B-splines on e_m}
\end{equation}
for $\mathbold{z} \in \mathbold{L}, m = 1,\dots,M$ and $k = 1,\dots,J_m$. Therefore, the B-splines $B_{m,k}$ are only supported on $\mathbold{e}_m$ and we denote with $B_\mathbold{e} = \lbrace B_{m,1},\dots,B_{m,{J_m}} \mid m = 1,\dots ,M\rbrace$ the set of all these B-splines. We further require that $J_m \geq 1$ for all $m = 1,\dots,M$ which is fulfilled if $\delta_m \leq \frac{d_m}{2}$ for all $m$.

\begin{figure}[t]
\center
\includegraphics[width=0.7\textwidth]{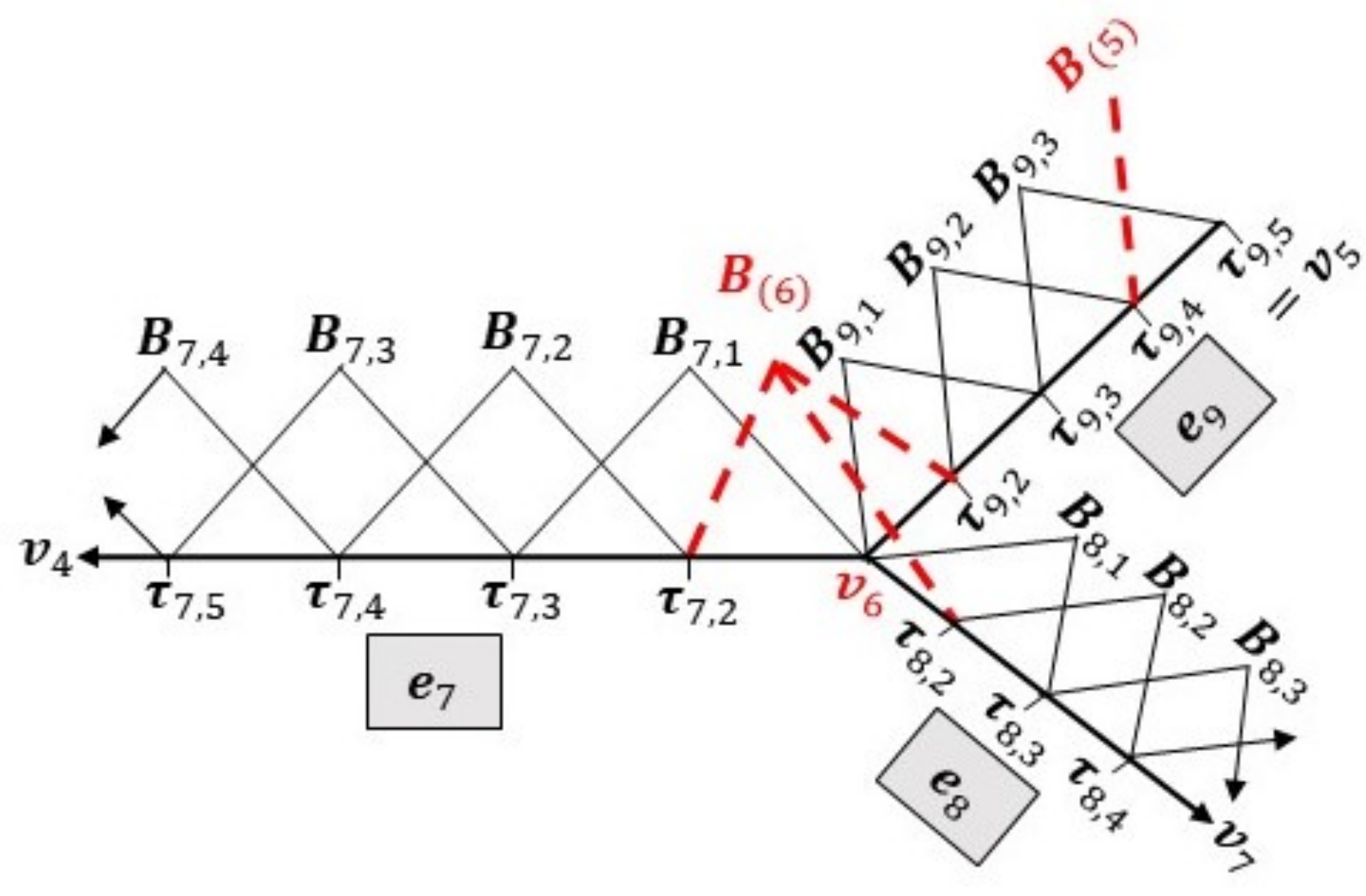} \hspace{2cm}
\caption{Schematic representation of a linear B-spline basis around $\mathbold{v}_6$ of Figure \ref{fig: example notation}. Here, $\mathbold{v}_6 = \mathbold{\tau}_{7,1} = \mathbold{\tau}_{8,1} = \mathbold{\tau}_{9,1}$ with adjacent segments $\mathbold{e}_7,\mathbold{e}_8,\mathbold{e}_9$. The red dotted lines show the linear B-splines which are contained in $B_\mathbold{v}$. The peaks of all B-splines are equal to 1 as in the Euclidean setting.}
\label{fig: linear B-splines}
\end{figure}

In addition to the B-splines defined by \eqref{eq: B-splines on e_m} we construct a single B-spline around each vertex $\mathbold{v}_i \in \mathbold{V}$. Therefore, we consider the $\text{deg}(\mathbold{v}_i)$ segments which have an endpoint equal to $\mathbold{v_i}$ and we numerate them (without loss of generality) with $\mathbold{e}_{1}, \dots, \mathbold{e}_{\text{deg}(\mathbold{v}_i)}$. Again, without loss of generality, let $ \mathbold{\tau}_{m_{1,1}} = \mathbold{v_i}, \dots, \mathbold{\tau}_{\text{deg}(\mathbold{v}_i),1} = \mathbold{v_i}$, i.e. we order the knots such that the first knot of every segment starting in $\mathbold{v}_i$ equals $\mathbold{v_i}$, see Figure \ref{fig: linear B-splines} as example. Then, we define the vertex specific B-spline $B_{(i)}$ for vertex $\mathbold{v}_i$ by
\begin{equation}
B_{(i)}(\mathbold{z}) = \sum_{k=1}^{\text{deg}(\mathbold{v}_i)}  \left[1-\frac{d_\mathbold{L}(\mathbold{v}_i, \mathbold{z})}{\delta_{k}}\right) \mathbbm{1}_{\left[\mathbold{v}_i; \mathbold{\tau}_{k,{2}}\right)}(\mathbold{z}).
\label{eq: definition B_v}
\end{equation}
for $\mathbold{z} \in \mathbold{L}$ and $i = 1,\dots,W$. These B-splines have support $\text{supp}(B_{(i)}) = \bigcup_{k=1}^{\text{deg}(\mathbold{v}_i)} [\mathbold{v}_i; \mathbold{\tau}_{k,{2}}) \subset \mathbold{L}$, i.e. they are supported on $\text{deg}(\mathbold{v}_i)$ segments. Note that all summands in \eqref{eq: definition B_v} are nonnegative and at most one of the summands is positive. This set of B-splines is denoted with $B_\mathbold{v} = \lbrace B_{(1)},\dots,B_{(W)} \rbrace$. Altogether, we specify the linear B-spline basis on $\mathbold{L}$ by $B = B_\mathbold{e} \cup B_\mathbold{v}$ with dimension $J = |B| = \sum_{m=1}^M J_m + |\mathbold{V}|$. For simplicity of presentation, we index from now on the B-spline Basis by $1,\dots,J$ and by construction, it holds that $\sum_{j=1}^J B_j(\mathbold{z}) = 1$ for $\mathbold{z} \in L$. In Figure \ref{fig: linear B-splines}, we depict linear B-splines around the vertex $\mathbold{v}_6$ with $\text{deg}(\mathbold{v}_6) = 3$ of the network that is shown Figure \ref{fig: example notation}.

\subsection{Intensity Estimation on a Network}

Exploiting the preliminary work of this section, we can now easily adopt the density estimation approach proposed by \cite{eilers1996flexible} for univariate data. On our geometric network $\mathbold{L}$, we specify a bin width $h_m$ on every segment $\mathbold{e}_m$ and then divide $\mathbold{e}_m$ into $N_m = \frac{d_m}{h_m}$ bins of the same length such that $\mathbold{L}$ is partitioned into  $N = \sum_{m=1}^M \frac{d_m}{h_m}$ bins in total. As for the knot distances $\delta_m$ it is clear, that $h_m$ can not be same for all curve segments of $\mathbold{L}$. However, also the bin widths are chosen very small when performing intensity estimation with penalized splines. We therefore specify a small global bin width $h$ and define according to \eqref{eq: delta_m}
\begin{equation*}
h_m = \begin{cases}
d_m / \lfloor \frac{d_m}{h} \rfloor, \quad \frac{d_m}{h} - \lfloor \frac{d_m}{h} \rfloor < 0.5 \\
d_m / \lceil \frac{d_m}{h} \rceil, \quad \frac{d_m}{h} - \lfloor \frac{d_m}{h} \rfloor \geq 0.5
\end{cases}.
\end{equation*}  
If the the left endpoint of $\mathbold{e}_m$ is $\mathbold{v}$, the bins are given by the $N_m$ subsets $[\mathbold{b}_{m,{k-1}}; \mathbold{b}_{m,k}) \subset \mathbold{e}_m$ for $k = 1,\dots,N_m$, where  $\mathbold{b}_{m,k} \in \mathbold{e}_m$ satisfies $d_\mathbold{L}(\mathbold{v}, \mathbold{b}_{m,k}) = kh_m$ and $\mathbold{b}_{m,0} = \mathbold{v}$. Each bin is characterized by its midpoint $\mathbold{z}_{m,k}$ which satisfies $d_\mathbold{L}(\mathbold{b}_{m,{k-1}}, \mathbold{z}_{m,k}) = d_\mathbold{L}(\mathbold{z}_{m,k}, \mathbold{b}_{m,k})$. 

Assume now that data on $n$ independently observed points $\mathbold{x}_i$ of the point process on the geometric network have been observed, with $i=1, \ldots ,n$.  For identifiability reasons we assume that the observed points are not equal to the vertices of the network, i.e. each point lies on a single edge. We define with $y_{m,k} \in \mathbb{N}_0$ the number of observations which are contained in the $k$-th bin of the $m$-th segment, i.e. 
\begin{equation*}
y_{m,k} = \# \lbrace \mathbold{x}_i \in \mathbold{L} \mid \mathbold{x}_i \in  [\mathbold{b}_{m,{k-1}}; \mathbold{b}_{m,k}), i = 1,\dots,n \rbrace
\end{equation*}
 for $m = 1,\dots,M$ and $k = 1,\dots,N_m$. Based on our considerations for the point process $\mathcal{X}$ we assume a Poisson distribution for the counts $y_{m,k}$ such that we have
\begin{equation}
y_{m,k} \mid \mathbold{z}_{m,k} \overset{\text{indep.}}{\sim} \text{Poi}(\lambda_{m,k}),
\label{eq: Poisson}
\end{equation}
where $\lambda_{m,k}$ is approximated through 
\begin{equation}
\lambda_{m,k} = \varphi_\mathcal{X}(\mathbold{z}_{m,k}) \cdot h_m = \exp \left( \nu_\mathcal{X}(\mathbold{z}_{m,k}) + \log h_m  \right).
\label{eq: lambda}
\end{equation}
We can consider $\log h_m$ as offset and aim to estimate $\nu_\mathcal{X}(\mathbold{z})$ as continuous log-intensity for $\mathbold{z} \in \mathbold{L}$ treating the pairs $(y_{m,k}, \mathbold{z}_{m,k})$ as independent observations from \eqref{eq: Poisson}. Therefore, we replace $\nu_\mathcal{X}(\mathbold{z})$ through the B-spline basis representation
\begin{equation}
\nu_\mathcal{X}(\mathbold{z}) = \sum_{j=1}^J B_j(\mathbold{z}) \gamma_j = \mathbold{B}(\mathbold{z}) \mathbold{\gamma},
\label{eq: row vector B}
\end{equation}
where $\mathbold{B}(\mathbold{z}) = (B_1(\mathbold{z}), \dots, B_J(\mathbold{z}))$ is a row vector consisting of the B-spline basis evaluated at $\mathbold{z} \in \mathbold{L}$ and $\mathbold{\gamma} = (\gamma_1,\dots,\gamma_J)^\top$ is the vector of B-spline coefficients that needs to be estimated from the data $\mathbold{x}_1,\dots,\mathbold{x}_n$. Imposing a penalty on the resulting Poisson likelihood leads to the penalized log-likelihood (constant terms are ignored)
\begin{align}
\ell_\mathcal{P}(\mathbold{\gamma}; \rho) = \sum_{m=1}^M \sum_{k=1}^{N_m} \left[ y_{m,k} \mathbold{B}(\mathbold{z}_{m,k}) \mathbold{\gamma} - \exp\left( \mathbold{B}(\mathbold{z}_{m,k}) \mathbold{\gamma} + \log h_m \right) \right] - \rho \mathcal{P}_r(\mathbold{\gamma}),
\label{eq: loglik}
\end{align}
where $\mathcal{P}_r(\mathbold{\gamma})$ is a penalty which is defined in the next section and $\rho$ is the smoothing parameter. It is also shown later how this smoothing parameter can be estimated.

If we replace $\mathbold{\gamma}$ in \eqref{eq: row vector B} with the maximum-likelihood estimate $\widehat{\mathbold{\gamma}} = \text{argmax}_\mathbold{\gamma} \ell_\mathcal{P}(\mathbold{\gamma}; \rho)$, then $\widehat{\nu}_\mathcal{X}(\mathbold{z}) = \mathbold{B}(\mathbold{z})\widehat{\mathbold{\gamma}}$ is an estimate of the log-intensity and thus $\widehat{\varphi}_\mathcal{X}(\mathbold{z}) = \exp \left( \widehat{\nu}_\mathcal{X}(\mathbold{z}) \right)$ is an estimate of the intensity of the point process $\mathcal{X}$ for $\mathbold{z} \in \mathbold{L}$. Also note that for a given $n$ an estimate of the density of $\mathcal{X}$ is given by $\widehat{f}_\mathcal{X}(\mathbold{z}) = \widehat{\varphi}_\mathcal{X}(\mathbold{z})/n.$

\subsection{Penalties on a Network}
In order to control smoothness of the intensity estimate and to overcome singularity issues, the penalty $\mathcal{P}_r(\mathbold{\gamma})$ multiplied with the smoothing parameter $\rho$ is subtracted from the maximum likelihood  criterion, which leads to the penalized log-likelihood \eqref{eq: loglik}. \cite{eilers1996flexible} proposed to impose a penalty on the vector of coefficients $\mathbold{\gamma}$ that is proportional to the $r$-th order differences of adjacent spline coefficients. The penalty is given by $\sum_{j=r+1}^J (\Delta^r \gamma_j)^2$ and for $r=1$ we have $\Delta^1 \gamma_j = \gamma_j - \gamma_{j-1}$. Higher order penalty terms can be calculated recursively using $\Delta^r(\gamma_j) = \Delta^1\Delta^{r-1}\gamma_j$ starting with the first order differences $\Delta^1\gamma_j$. It is straightforward to extend this idea to penalties on a network. Let $i,j = 1,\dots,J$ where $J$ is the dimension of the B-spline basis on the geometric network. According to the one-dimensional case, we are interested in the set of pairwise adjacent B-splines or their coefficients, respectively. Hence, we can view the B-splines on the geometric network $\mathbold{L}$ itself as a network graph $L_B$ which is defined through a $J \times J$ adjacency matrix $\mathbold{A}$. From the definition of the linear B-splines, it follows that $\mathbold{A}(i,j) = 1$, if $\text{supp}(B_i) \cap \text{supp}(B_j) \neq \emptyset$ and else $\mathbold{A}(i,j) = 0$. In order to define penalties of arbitrary order, we need the $J \times J$ shortest path matrix $\mathbold{S}_\mathbold{A}$ where $\mathbold{S}_\mathbold{A}(i,j) = s$, if the B-Splines $B_i$ and $B_j$ have minimum distance $s$ in $L_B$. This all-pairs shortest path problem can be solved with complexity $\mathcal{O}(J|\mathcal{A}|)$ where $|\mathbold{A}|$ is the number of non-zero entries in $\mathbold{A}$ \citep{chan2012all}. For illustration, consider again Figure \ref{fig: linear B-splines}. Here, $B_{7,1}$ is adjacent to $B_{7,2}$ as well as $B_{(6)}$ and the shortest path from $B_{7,2}$ to $B_{8,1}$ via $B_{7,1}$ and $B_{(6)}$ has length $3$ in $L_B$.

Now, let $\mathcal{D}_1 = \lbrace (i, j) \mid  \mathbold{S}_\mathbold{A}(i,j) = 1, 1\leq i < j \leq J \rbrace$. According to \cite{eilers1996flexible} we penalize neighboring coefficients. A first order penalty is then defined by 
\begin{equation}
\mathcal{P}_1(\mathbold{\gamma}) =  \sum_{\mathcal{D}_1} (\gamma_i - \gamma_j)^2 = (\mathbold{D}_1\mathbold{\gamma})^\top (\mathbold{D}_1 \mathbold{\gamma}) = \mathbold{\gamma}^\top \mathbold{K}_1 \mathbold{\gamma},
\label{eq: first order penalty}
\end{equation} 
where $\mathbold{D}_1 \in \mathbb{Z}^{|\mathcal{D}_1| \times J}$ and $\mathbold{K}_1 = \mathbold{D}_1^\top \mathbold{D}_1 \in \mathbb{Z}^{J \times J}$ define the difference matrix and the resulting quadratic form according to the pairwise differences in \eqref{eq: first order penalty}. Further, let 
\begin{equation*}
\mathcal{D}_2 = \lbrace (i,k,j) \mid \mathbold{S}_\mathbold{A}(i,j) = 2, \mathbold{S}_\mathbold{A}(i,k) = \mathbold{S}_\mathbold{A}(k,j) = 1, 1\leq i < j \leq J   \rbrace.
\end{equation*}
Therewith, a second order penalty can be defined by 
\begin{equation}
\mathcal{P}_2(\mathbold{\gamma}) = \ \sum_{\mathcal{D}_2} ((\gamma_i - \gamma_k)-(\gamma_k - \gamma_j))^2 = \ \sum_{\mathcal{D}_2} (\gamma_i -2\gamma_k + \gamma_j)^2 = (\mathbold{D}_2\mathbold{\gamma})^\top (\mathbold{D}_2 \mathbold{\gamma}) = \mathbold{\gamma}^\top \mathbold{K}_2 \mathbold{\gamma},
\label{eq: second order penalty}
\end{equation}
where $\mathbold{D}_2 \in \mathbb{Z}^{|\mathcal{D}_2| \times J}$ and again, $\mathbold{K}_2 = \mathbold{D}_2^\top \mathbold{D}_2 \in \mathbb{Z}^{J \times J}$ results as matrix version from the sum in \eqref{eq: second order penalty}. For illustration, we revisit the B-splines which we depict in Figure \ref{fig: linear B-splines}, but, restricted to the B-splines $B_1 = B_{7,2}, B_2 = B_{7,1}, B_3 = B_{(6)}, B_4 = B_{8,1}$ and $B_5 = B_{9,1}$. Thus, for $\mathbold{\gamma} = (\gamma_1,\dots,\gamma_5)$ the first- and second order penalties $\mathcal{P}_1(\mathbold{\gamma})$ and $\mathcal{P}_2(\mathbold{\gamma})$ are defined by the difference matrices
\begin{equation*}
\mathbold{D}_1 = \begin{pmatrix}
1 & -1 & 0 & 0 & 0 \\
0 & 1 & -1 & 0 & 0 \\
0 & 0 & 1 & -1 & 0 \\
0 & 0 & 1 & 0 & -1
\end{pmatrix} \quad  \text{and} \quad
\mathbold{D}_2 = \begin{pmatrix}
1 & -2 & 1 & 0 & 0 \\
0 & 1 & -2 & 1 & 0 \\
0 & 1 & -2 & 0 & 1 \\
0 & 0 & -2 & 1 & 1
\end{pmatrix},
\end{equation*}
respectively. Exploiting the shortest path matrix $\mathbold{S}_\mathbold{A}$ we can define penalties of any order $r$, but usually first and second order differences are used when applying penalized splines. 

\subsection{Estimation of the Smoothing Parameter}

For the estimation of the smoothing parameter $\rho$, we apply the generalized Fellner-Schall method \citep{wood2017generalized} which is an iterative procedure to estimate the smoothing parameter in generalized additive models \citep{wood2017generalized2}. The idea behind the Fellner-Schall method is to apply a mixed model approach and to compute the log Laplace approximate marginal likelihood of the model and its derivative with respect to the smoothing parameter $\rho$. In each iteration step we estimate the model parameters $\widehat{\mathbold{\gamma}}_\rho$ by maximizing the penalized log-likelihood \eqref{eq: loglik} while taking $\rho$ from the previous cycle. Then, the update $\rho_\text{new}$ is calculated through
\begin{equation}
\rho_\text{new} = \rho \frac{\text{tr}((\rho \mathbold{K}_r)^- \mathbold{K}_r) - \text{tr}((\mathbold{B}^\top \mathbold{W}(\widehat{\mathbold{\gamma}}_\rho) \mathbold{B} + \rho \mathbold{K}_r)^{-1}\mathbold{K}_r)}{\widehat{\mathbold{\gamma}}_\rho^\top \mathbold{K}_r \widehat{\mathbold{\gamma}}_\rho}
\label{eq: update rho}
\end{equation} 
where $\text{tr}(\cdot)$ denotes the trace operator for diagonal matrices and $(\rho \mathbold{K}_r)^-$ denotes a generalized inverse of $\rho \mathbold{K}_r$. The design matrix $\mathbold{B}\in \mathbb{R}^{N\times J}$ of the Poisson model is build by storing the row vectors $ \mathbold{B}(\mathbold{z}_{m,k})$, which are defined accordingly to \eqref{eq: row vector B} for $m = 1,\dots,M$ and $k = 1,\dots,N_m$, as a matrix. Furthermore, $\mathbold{W}(\widehat{\mathbold{\gamma}}_\rho) = \text{diag}(\widehat{\lambda}_{1,1}, \dots,\widehat{\lambda}_{1,N_1},\dots,\widehat{\lambda}_{M,1},\dots,\widehat{\lambda}_{M,N_M})$ is a weight matrix, where $\widehat{\lambda}_{m,k} = \exp \left( \widehat{\gamma}_\mathcal{X}(\mathbold{z}_{m,k}) + \log h_m \right)$ is defined through \eqref{eq: lambda}. The matrix $\mathbold{B}^\top \mathbold{W}(\widehat{\mathbold{\gamma}}_\rho) \mathbold{B} + \rho \mathbold{K}_r$ is the expected Hessian of our model and is therefore positive definite, which guarantees that $\rho_\text{new} > 0$ \citep{wood2017generalized}. The iterative procedure stops, if $\rho_\text{new}$ in \eqref{eq: update rho} differs only slightly from $\rho$.

\section{Simulation Study}
\label{sec: simulation}

\begin{figure}[!h]
\center
\includegraphics[width = 0.3\textwidth]{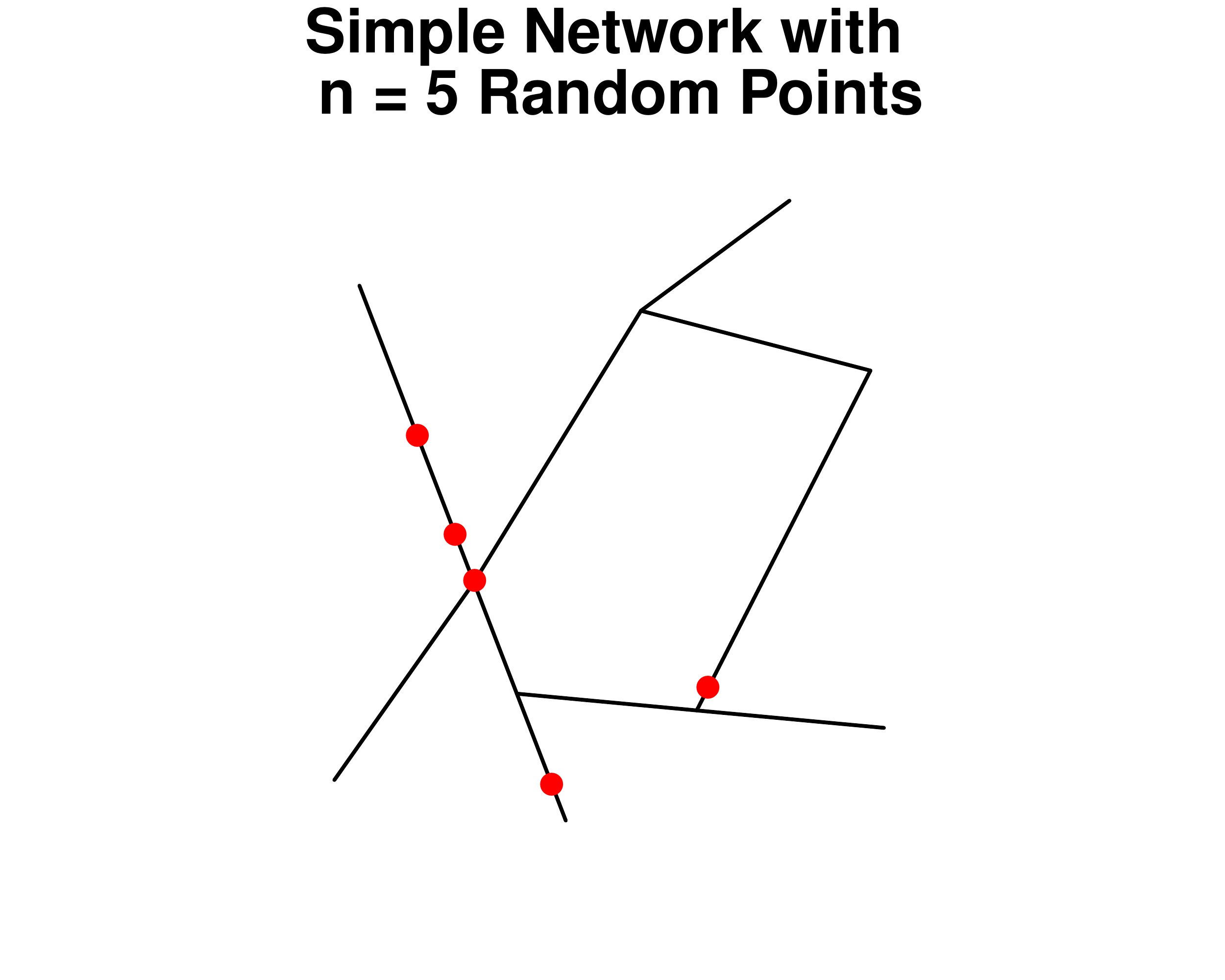} 
\includegraphics[width = 0.3\textwidth]{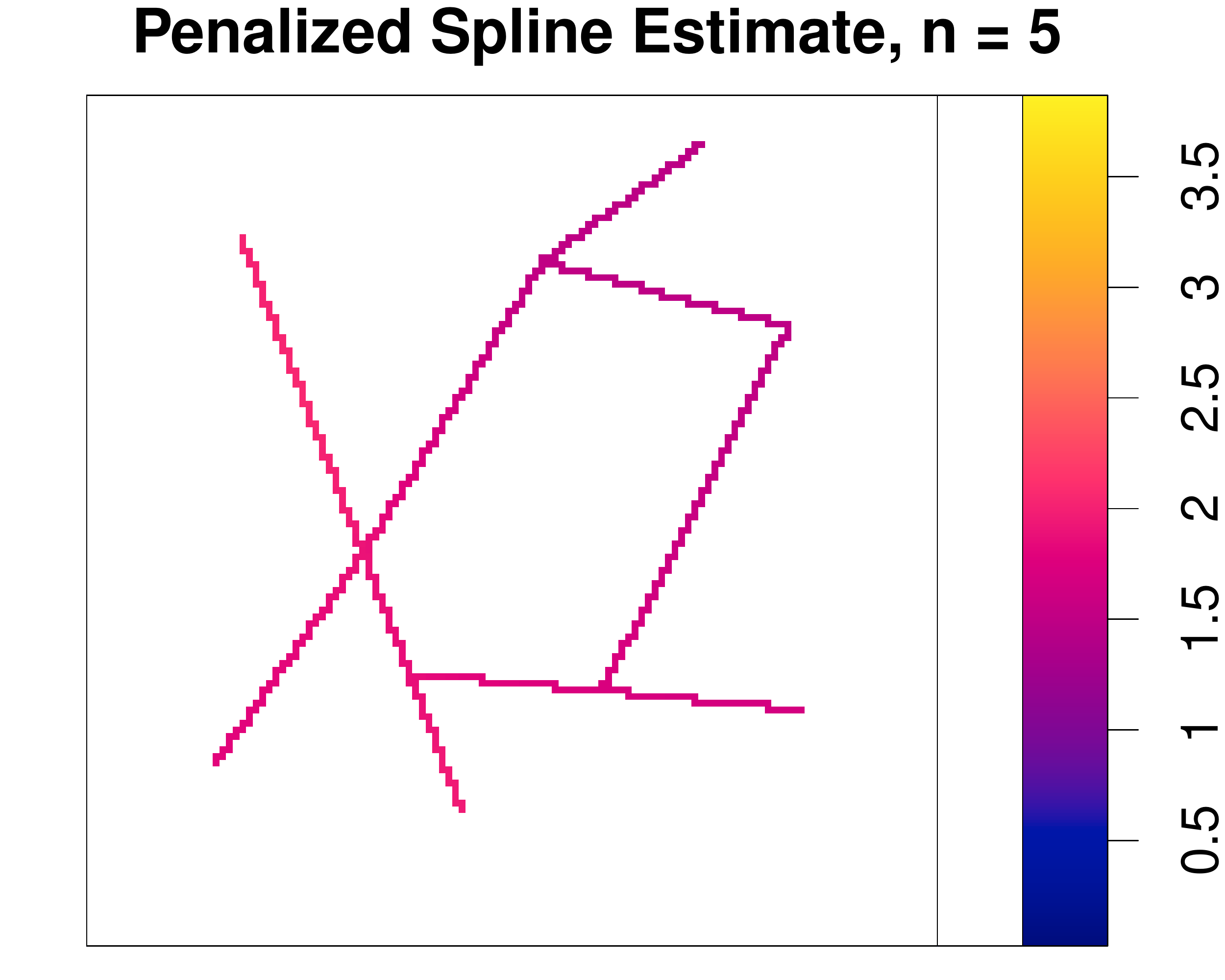}
\includegraphics[width = 0.3\textwidth]{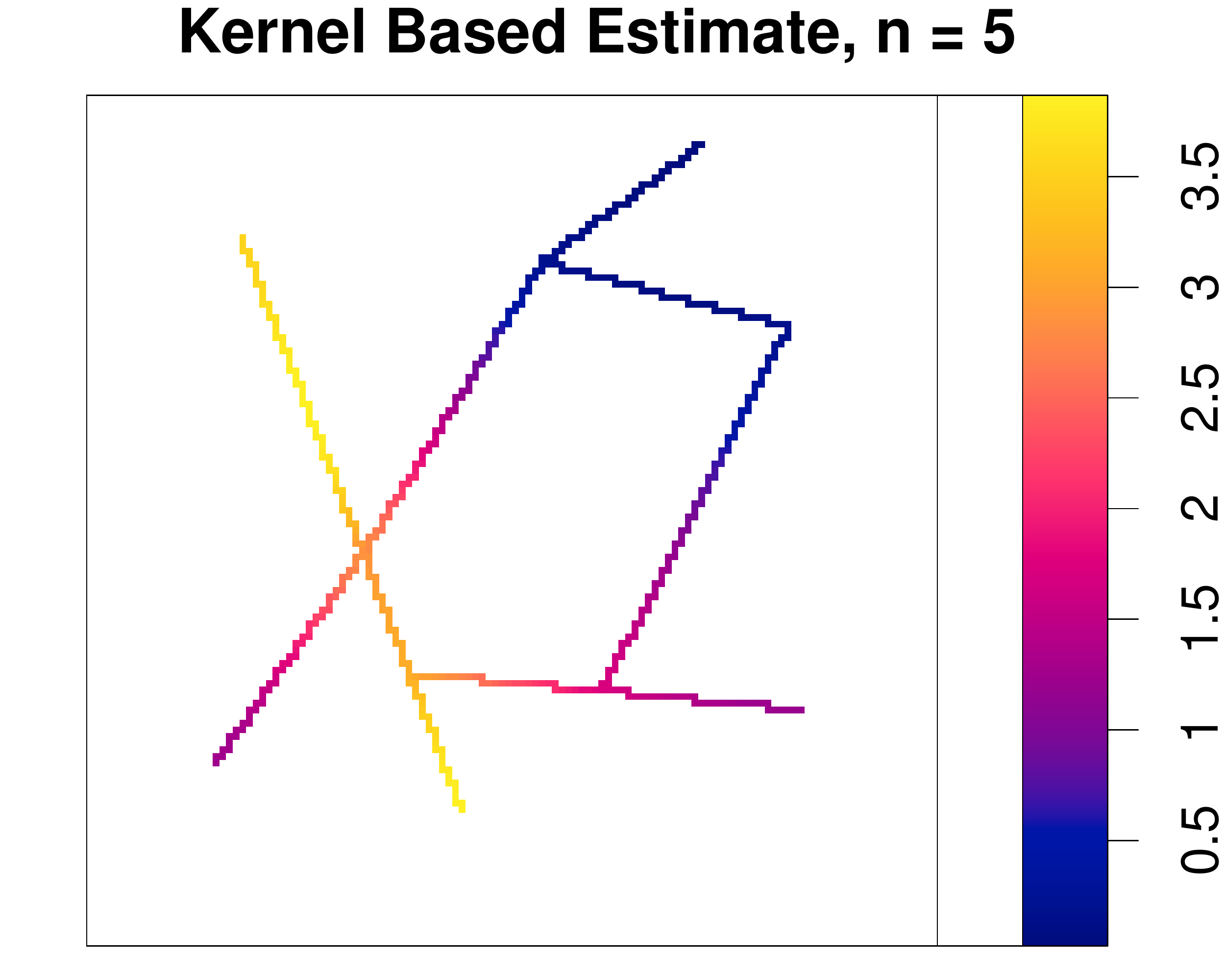}
\includegraphics[width = 0.3\textwidth]{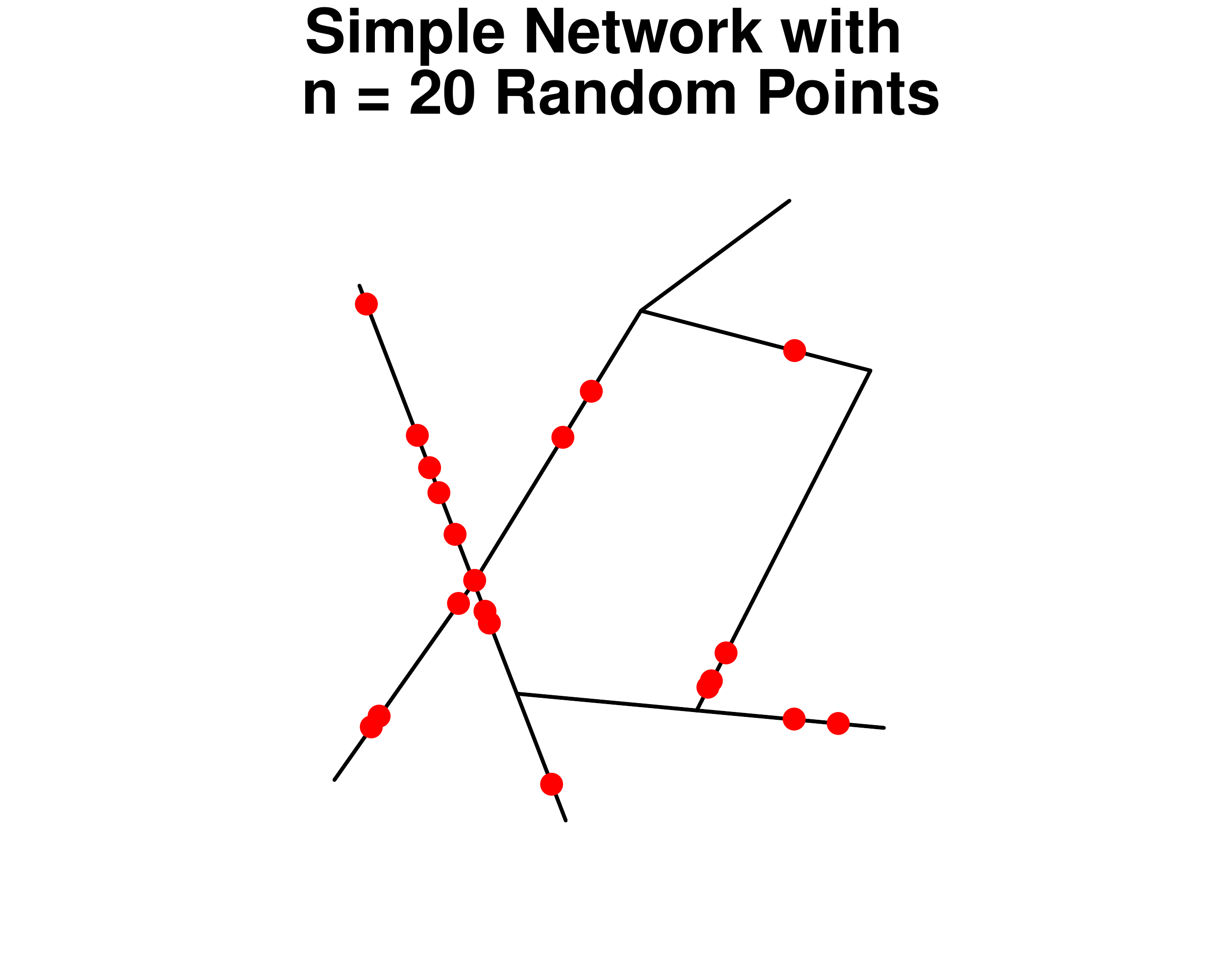} 
\includegraphics[width = 0.3\textwidth]{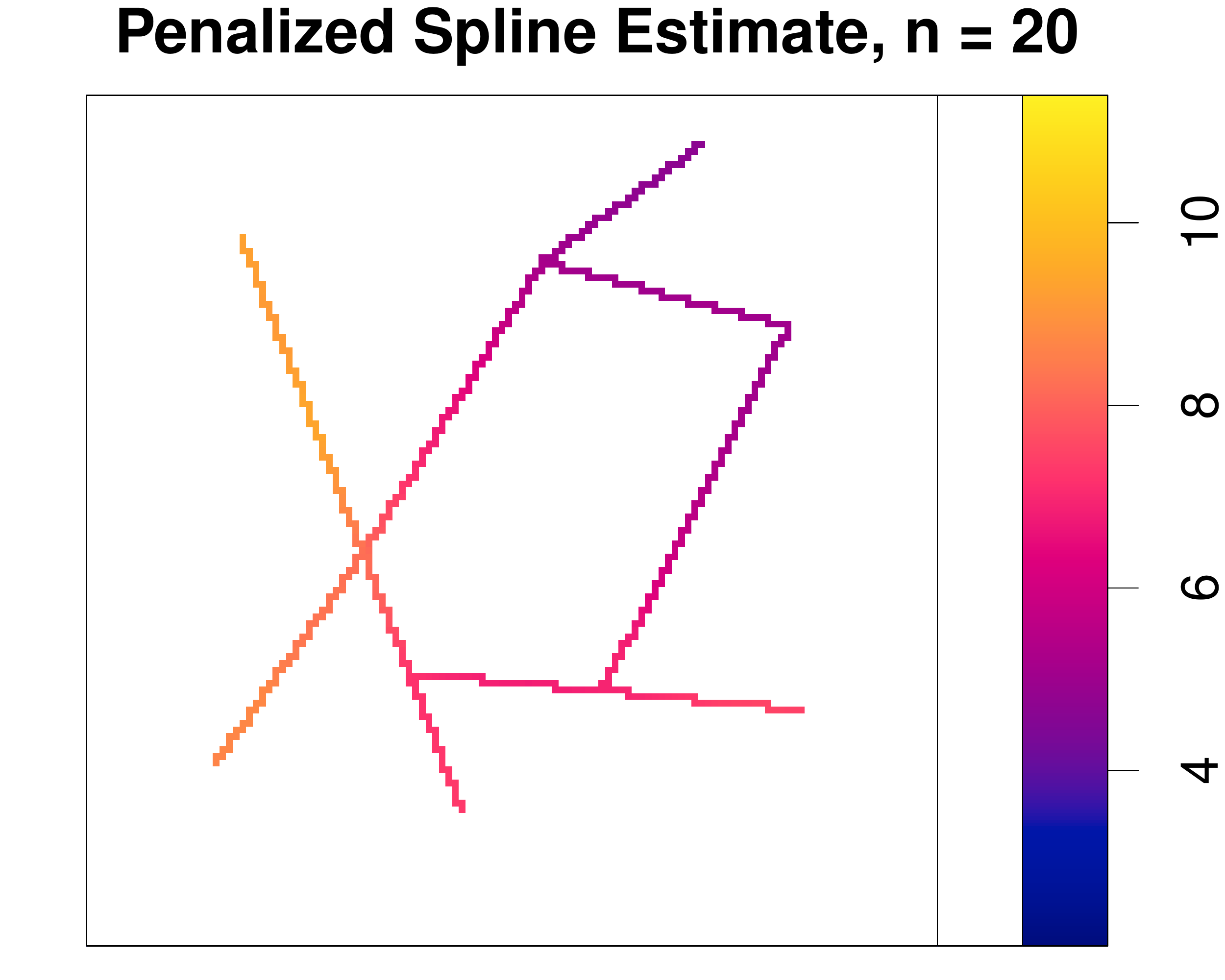}
\includegraphics[width = 0.3\textwidth]{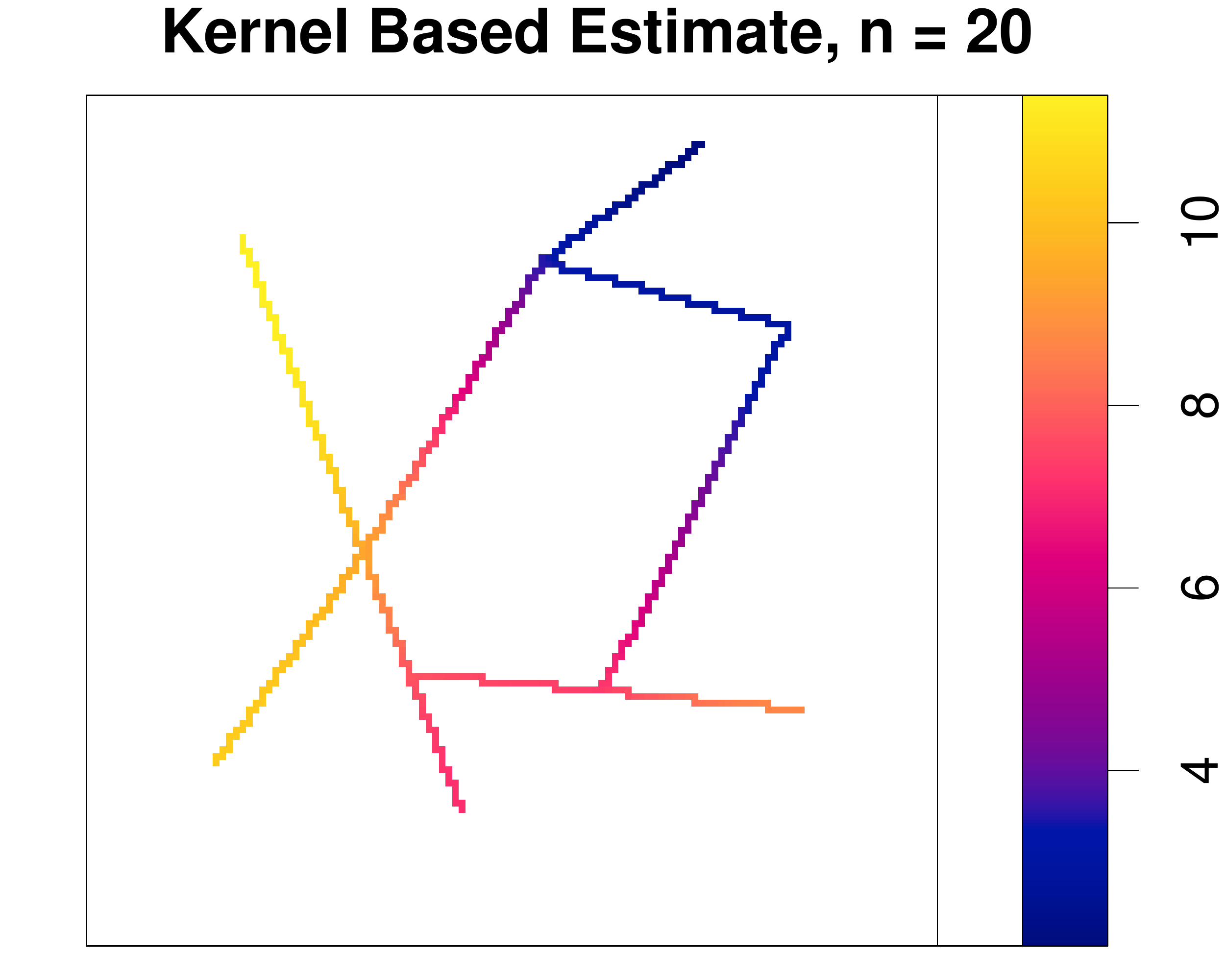}
\includegraphics[width = 0.3\textwidth]{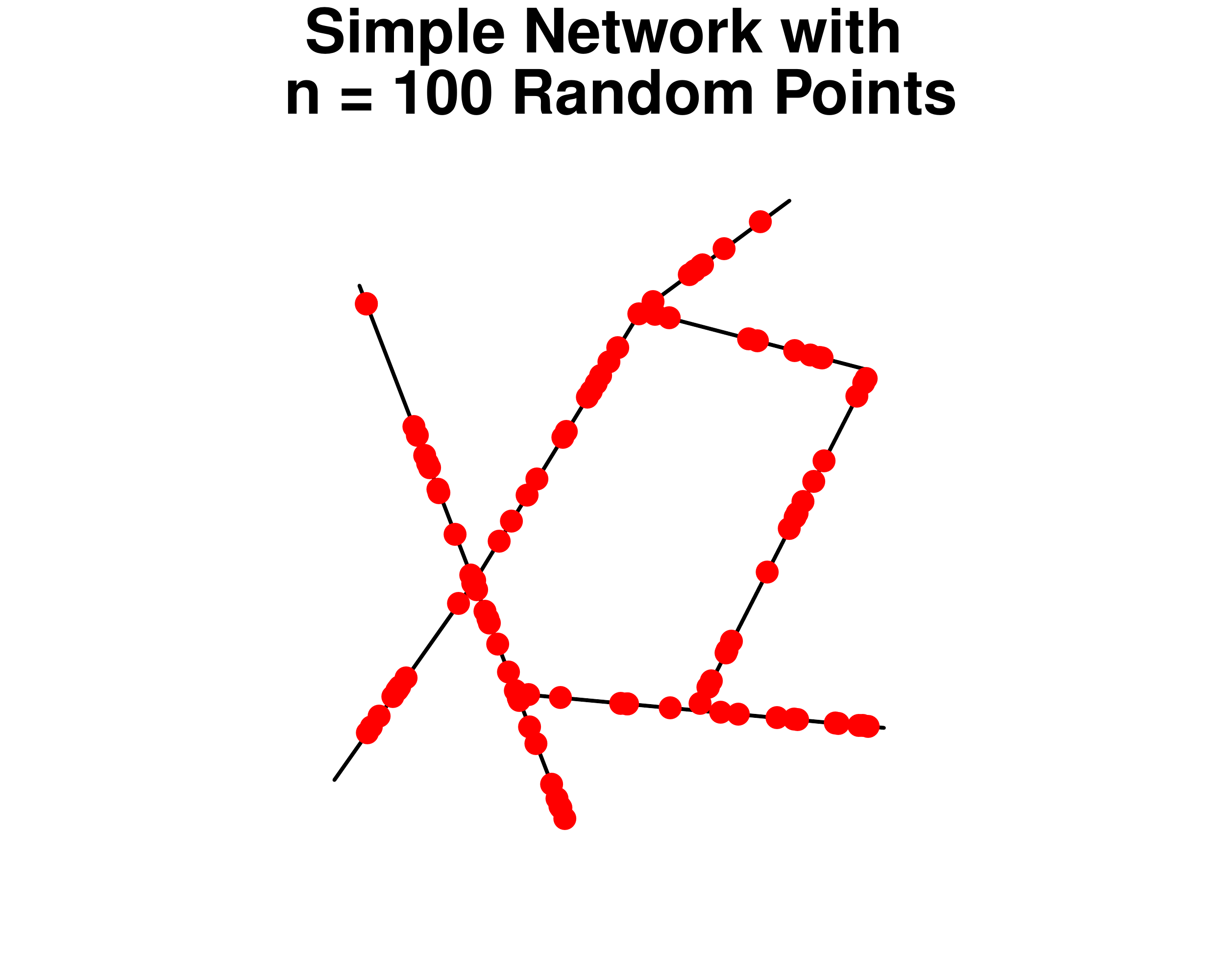} 
\includegraphics[width = 0.3\textwidth]{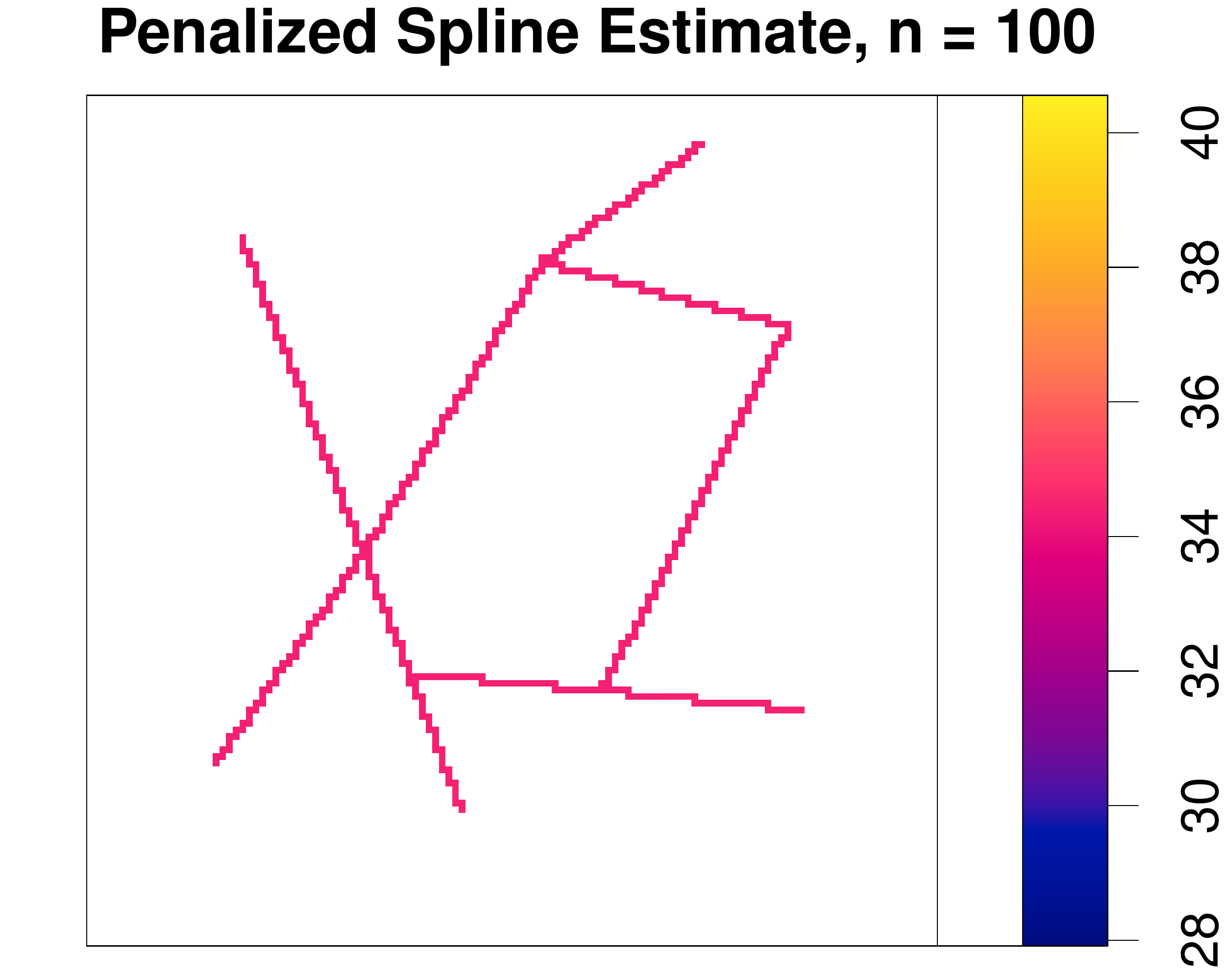}
\includegraphics[width = 0.3\textwidth]{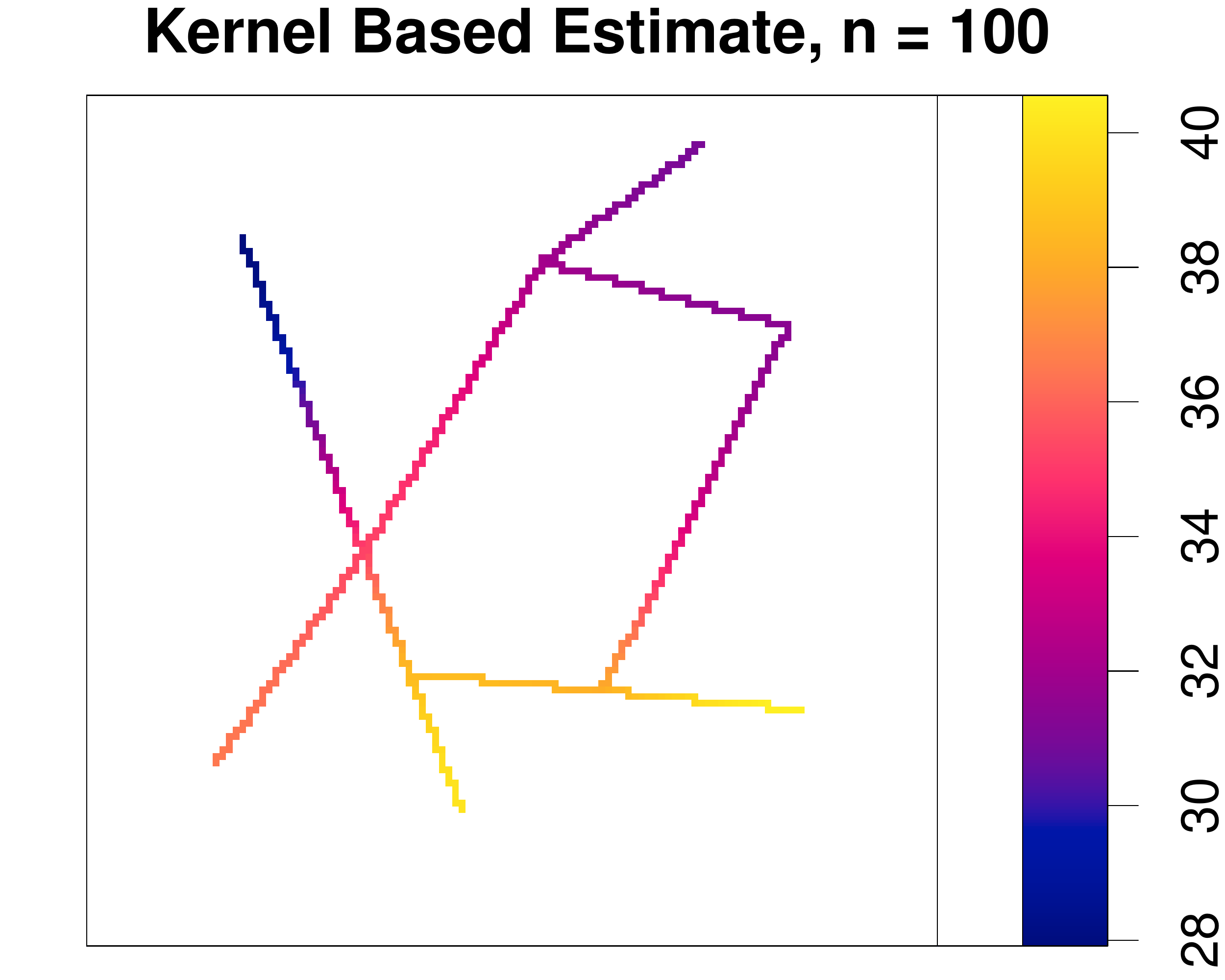}
\includegraphics[width = 0.3\textwidth]{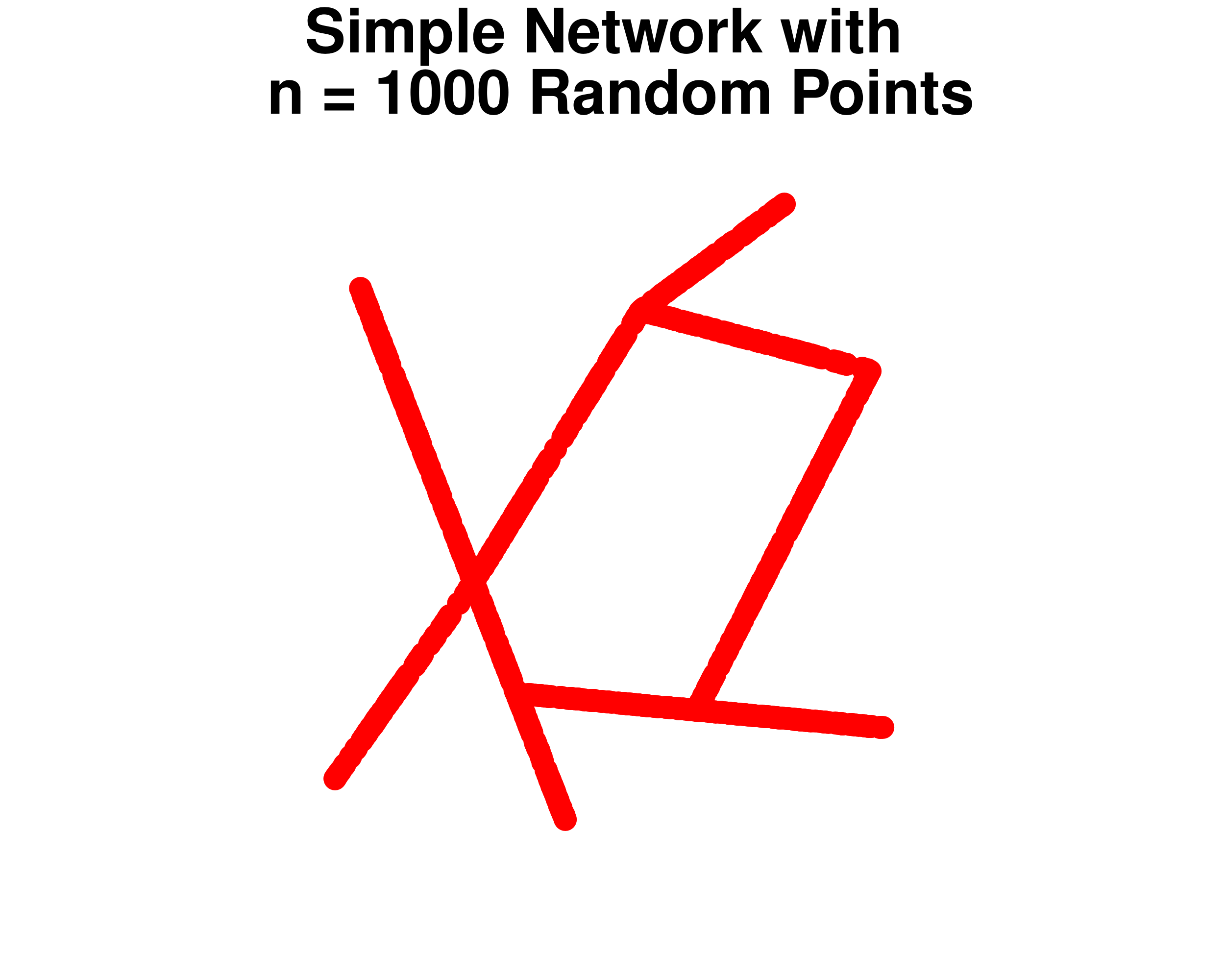} 
\includegraphics[width = 0.3\textwidth]{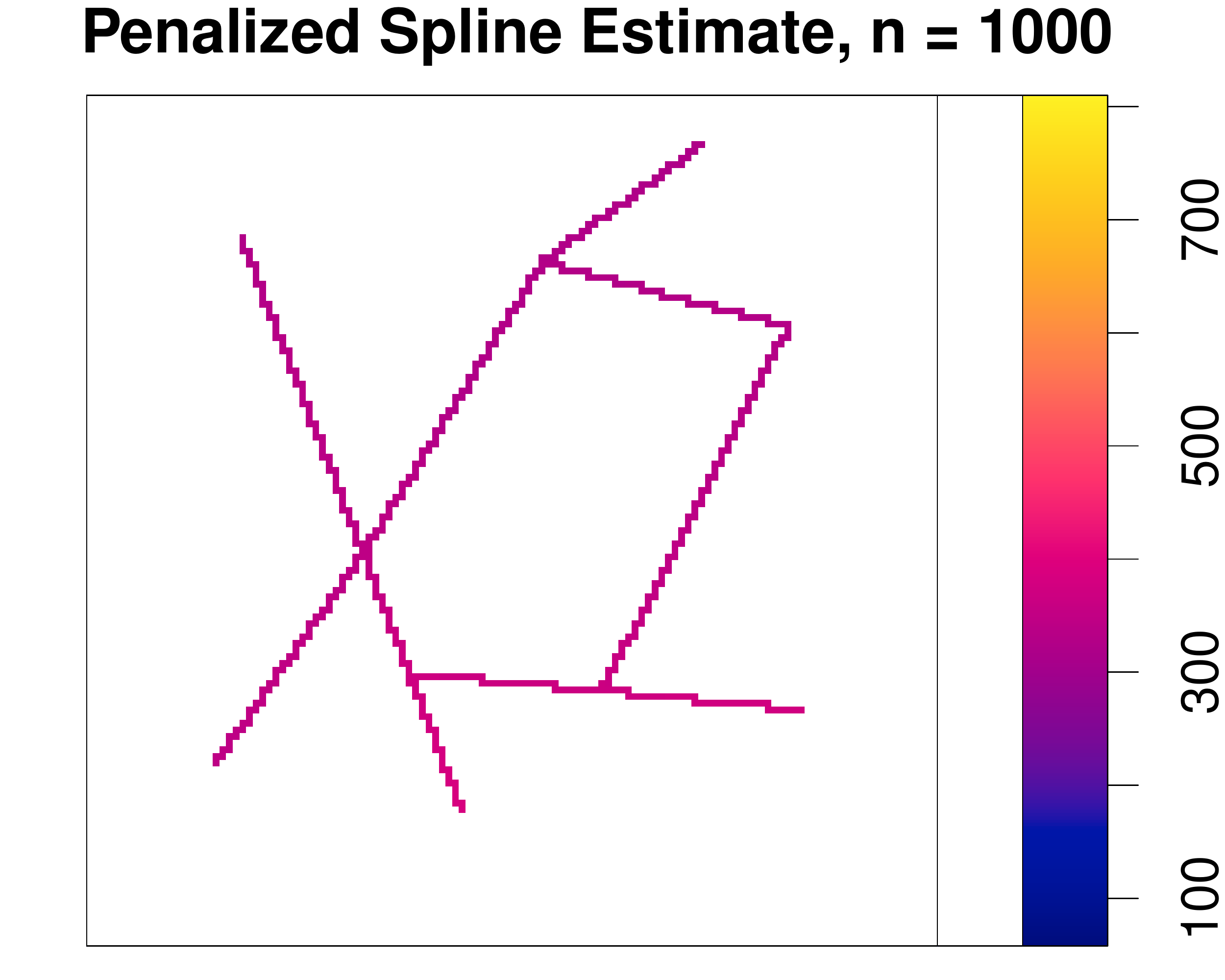}
\includegraphics[width = 0.3\textwidth]{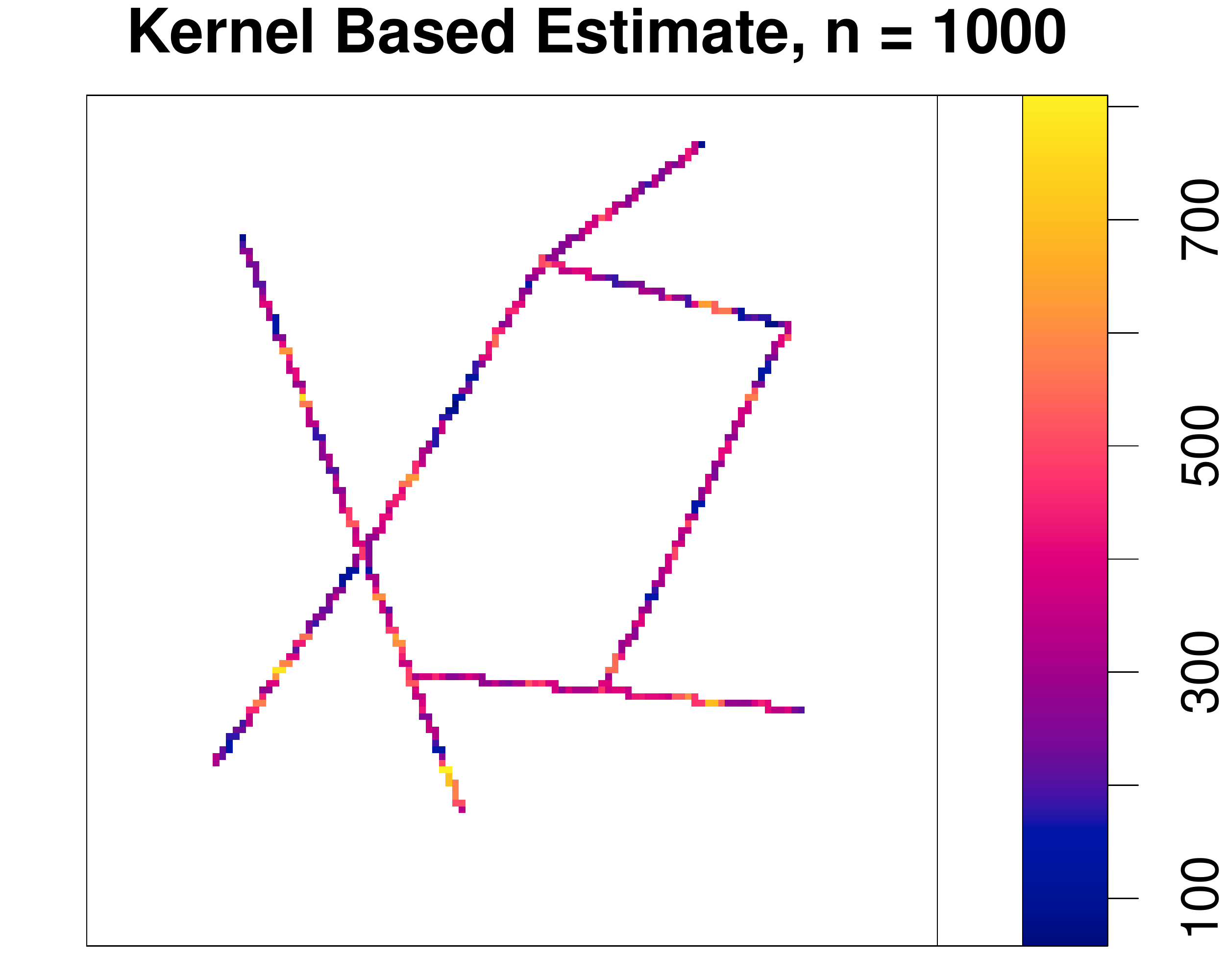}
\caption{Simple network with $n = 5, 20, 100, 1000$ uniformly distributed random points (left panels); Intensity estimates based on penalized splines  (middle panels) and kernel estimation (right panels).}
\label{fig: simplenet}
\end{figure}

In this section we exploit simulated data in order to explore the performance of penalized spline intensity estimation on geometric networks. The general setting of this study is that we consider the rather small network \texttt{simplenet} from the \texttt{spatstat}  package \citep{baddeley2015spatial} with $|\mathbold{E}| = |\mathbold{V}| = 10$ and total length $|\mathbold{L}| = 2.905$. On this network we specify an intensity function $\varphi_{\mathcal{X}_n}$, where $ \int_\mathbold{L} \varphi_{\mathcal{X}_n}(\mathbold{z}) \dif \mathbold{z} = n$. Therefore, we independently simulate $n$ random points according to $\varphi_{\mathcal{X}_n}$ and estimate the intensity of the simulated point process with penalized spline smoothing. This is opposed to a kernel based intensity estimate with automatic bandwidth selection as implemented in the \texttt{spatstat}  package. Our routine is also implemented in R\footnote{The R routines including the examples of Section \ref{sec: simulation} and Section \ref{sec: application} are available on \url{https://github.com/MarcSchneble/NetworkSplines}.} and heavily exploits the geometric network representation of the \texttt{spatstat} package. Here, a geometric network is embedded in the plane and consists merely of straight line segments, but we can align several line segments to a piecewise linear curve.

\subsection{Uniform Intensity -- Graphical Comparison with Kernel Based Estimation}

To begin with, we specify a uniform intensity on $\mathbold{L}$, i.e. $\varphi_{\mathcal{X}_n}(\mathbold{z}) = n/|\mathbold{L}|$ for all $\mathbold{z} \in \mathbold{L}$. We independently simulate random samples $\lbrace \mathbold{x}_1, \dots, \mathbold{x}_{n} \rbrace$ for $n = 5,20,100,1000$ according to $\varphi_{\mathcal{X}_n}$. The four exemplary simulated point processes on $\mathbold{L}$ are shown in the left panels of Figure \ref{fig: simplenet}. We estimate the intensity of these processes exploiting both, the penalized spline and the kernel based approach. For the former method, we use $\delta = 0.05, h = 0.01$ with a first-order penalty. 

When estimating the smoothing parameter with $n = 100$ we get $\widehat{\rho} \rightarrow \infty$ such that the penalty on the log-likelihood of the model affects $\widehat{\gamma}_1 \approx \dots \approx \widehat{\gamma}_P$. Therefore, when using the penalized spline method we estimate nearly a constant intensity, i.e. $\widehat{\varphi}_{\mathcal{X}_{100}}(\mathbold{z}) \approx 100/{|\mathbold{L}|} = 34.432$ for all $\mathbold{z} \in \mathbold{L}$, which can be observed in the middle panel of the third row in Figure \ref{fig: simplenet}. The optimal bandwidth $\widehat{\kappa}$ for $n = 100$ when using the kernel method is $\widehat{\kappa} = 0.276$. In contrast to our method, the intensity calculated therewith yields values for $\widehat{\varphi}_{\mathcal{X}_{100}}(\mathbold{z})$ within the interval $[28, 40]$, see the right panel of the third row in Figure \ref{fig: simplenet}. This originates from the locality of the kernel smoother.

A more extreme discrepancy between the two estimation methods is observable for $n = 1000$, see the bottom row of Figure \ref{fig: simplenet}. Again, our method estimates a nearly constant intensity whereas the kernel method yields a very rough estimate. Here, the bandwidth selection algorithm estimates an optimal bandwidth of $\widehat{\kappa} = 0.0063$, which is very small compared to the length of the single line segments. Furthermore, the top two rows of Figure \ref{fig: simplenet} show that even with a small sample size, we are still able to roughly estimate a constant intensity with our method and again, the kernel based method leads to a very high variation in the intensity estimate along the geometric network.

\begin{figure}[t]
\center
\includegraphics[width = 0.4 \textwidth]{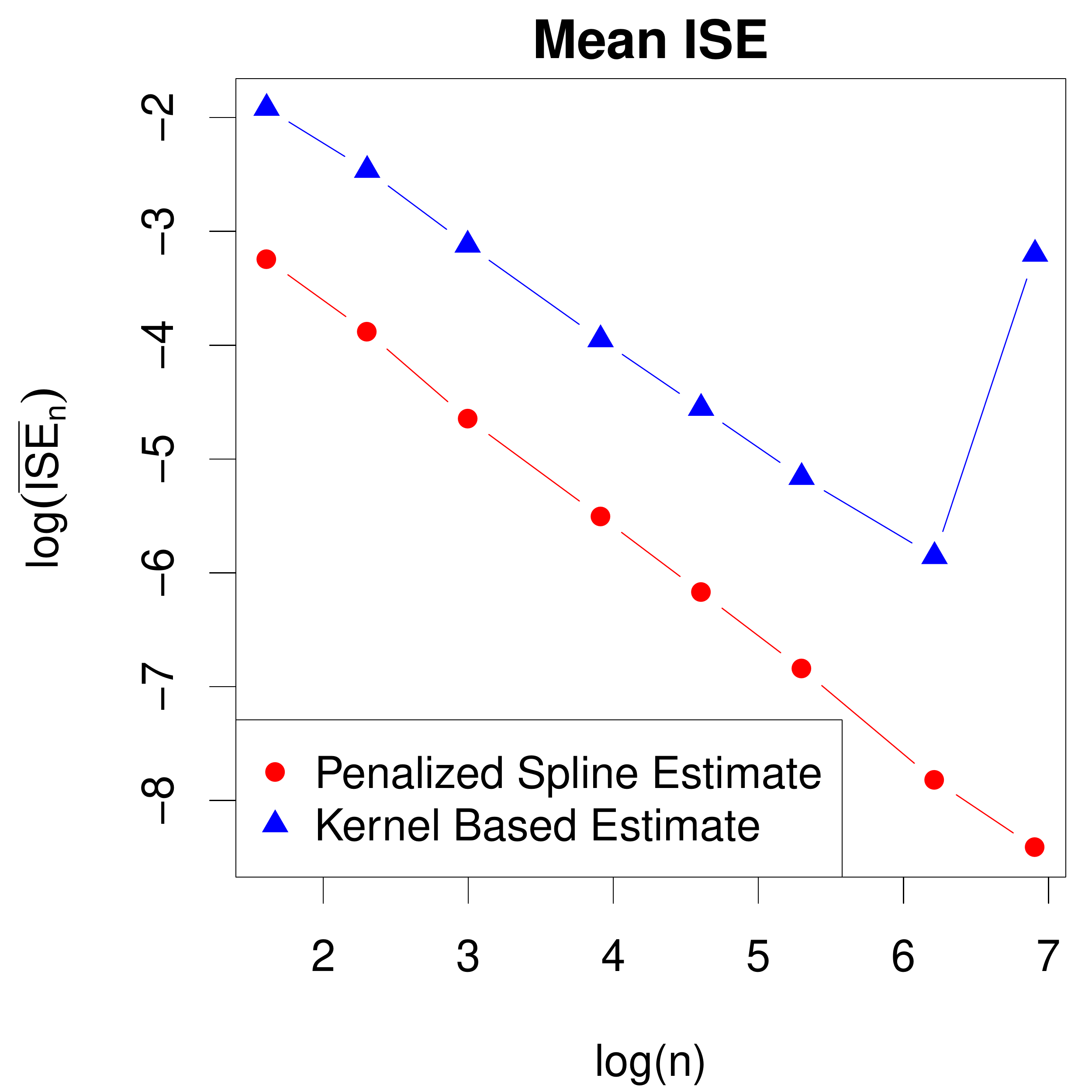} 
\includegraphics[width = 0.4 \textwidth]{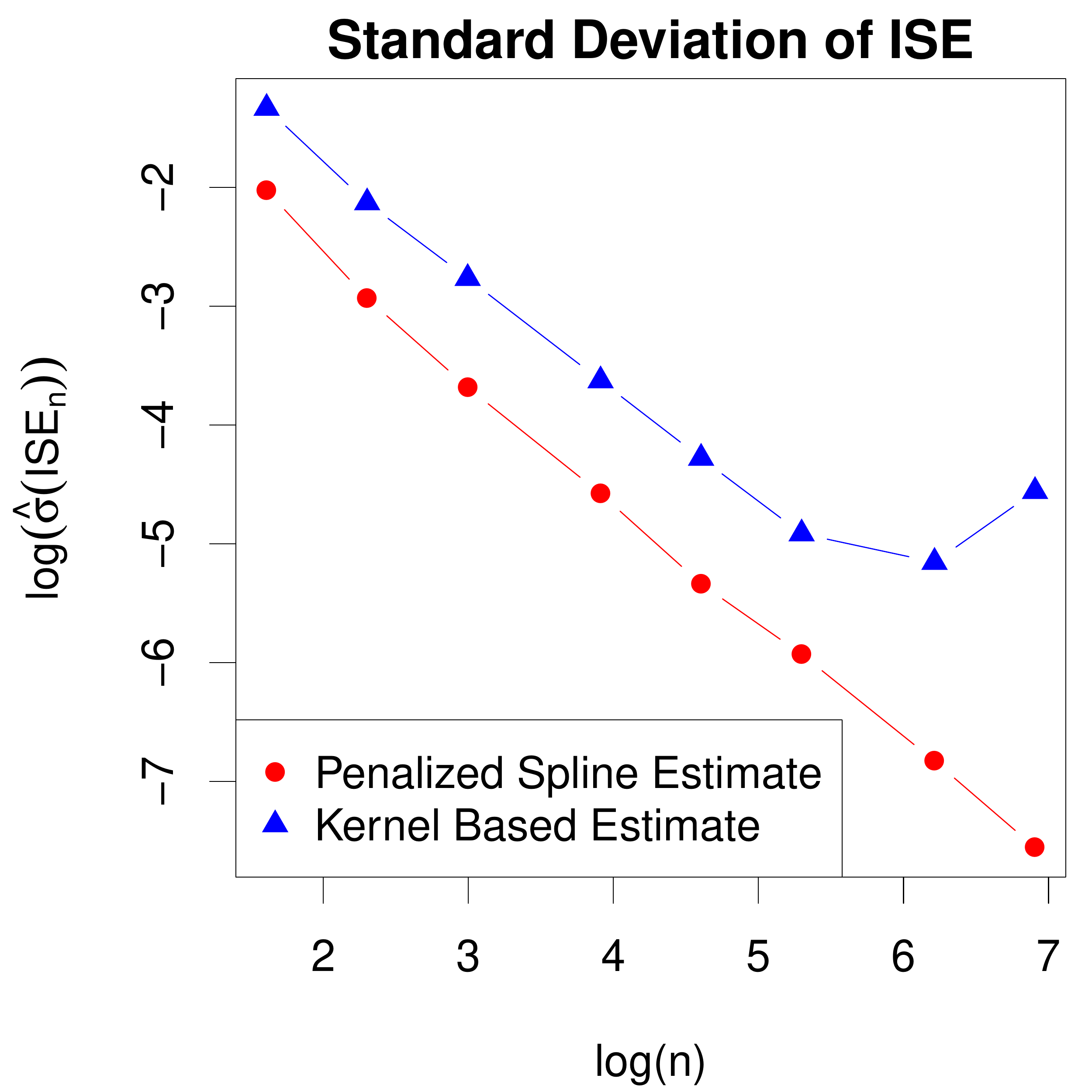} 
\caption{Mean and standard deviation of penalized spline based and kernel based $\text{ISE}$ for $n = 5,10,20,50,100,200,500,1000$ on a log-log-scale, where $\varphi_{\mathbold{X}_n}$ is defined according to a uniform intensity on $\mathbold{L}$.}
\label{fig: ISE uniform n}
\end{figure}

\begin{table}[t]
\center
\begin{tabular}{rllll}
\toprule
\textbf{Uniform Intensity} & $h = 0.1$  & $h = 0.05$ & $h = 0.01$ & $h = 0.005$ \\ 
 \midrule
 $\delta = 0.1$  & 1.65 (4.19)  & 2.24 (5.14) & 1.91 (4.72) & 2.24 (6.08) \\
 $\delta = 0.05$ &  --          & 2.08 (5.62) & 2.40 (5.71) & 2.15 (5.03) \\
 $\delta = 0.01$ &  --          &     --      & 1.97 (5.29) & 2.18 (5.57) \\
 \bottomrule
\end{tabular}
\caption{Mean (standard deviation) of $\text{ISE}$ if $\varphi_{\mathcal{X}_n}$ follows a uniform intensity on $\mathbold{L}$ with $n = 100$ as well as different choices of $\delta$ and $h$. All values in this table are multiplied by factor 1,000.}
\label{tab: ISE uniform delta h} 
\end{table}

\subsection{Uniform Intensity -- Integrated Squared Error}

\begin{figure}[t]
\center
\includegraphics[width = 0.4 \textwidth]{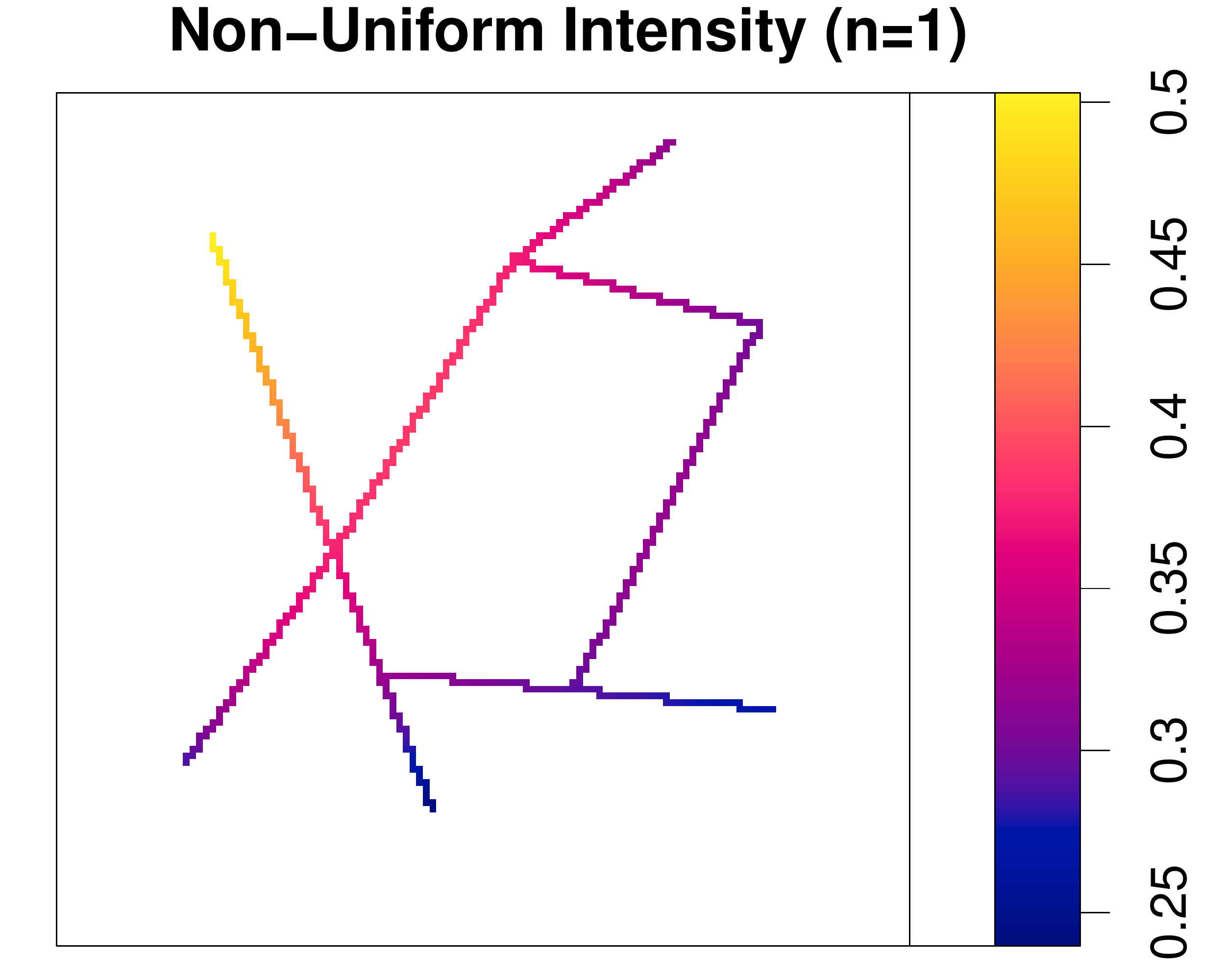} \\
\includegraphics[width = 0.4 \textwidth]{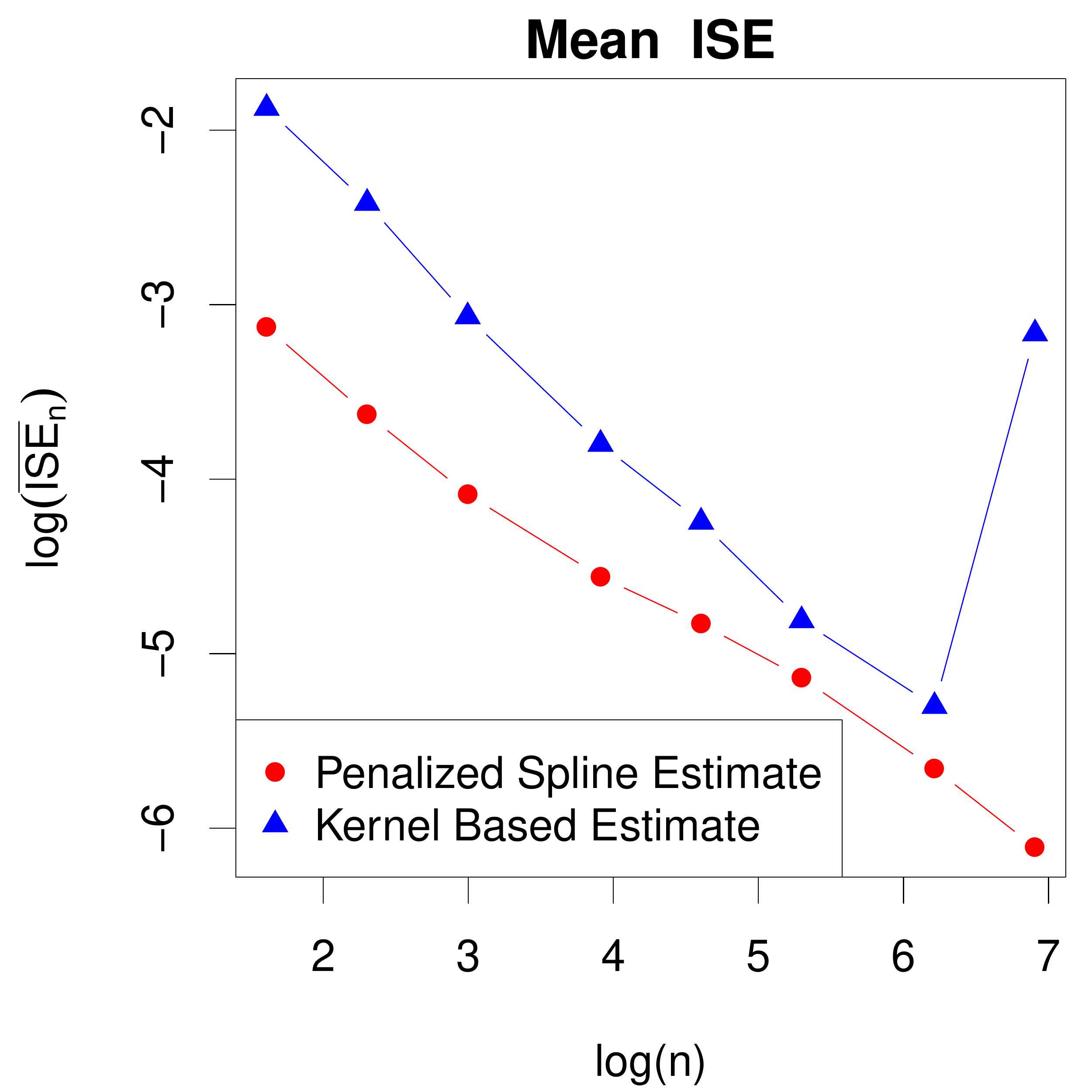} 
\includegraphics[width = 0.4 \textwidth]{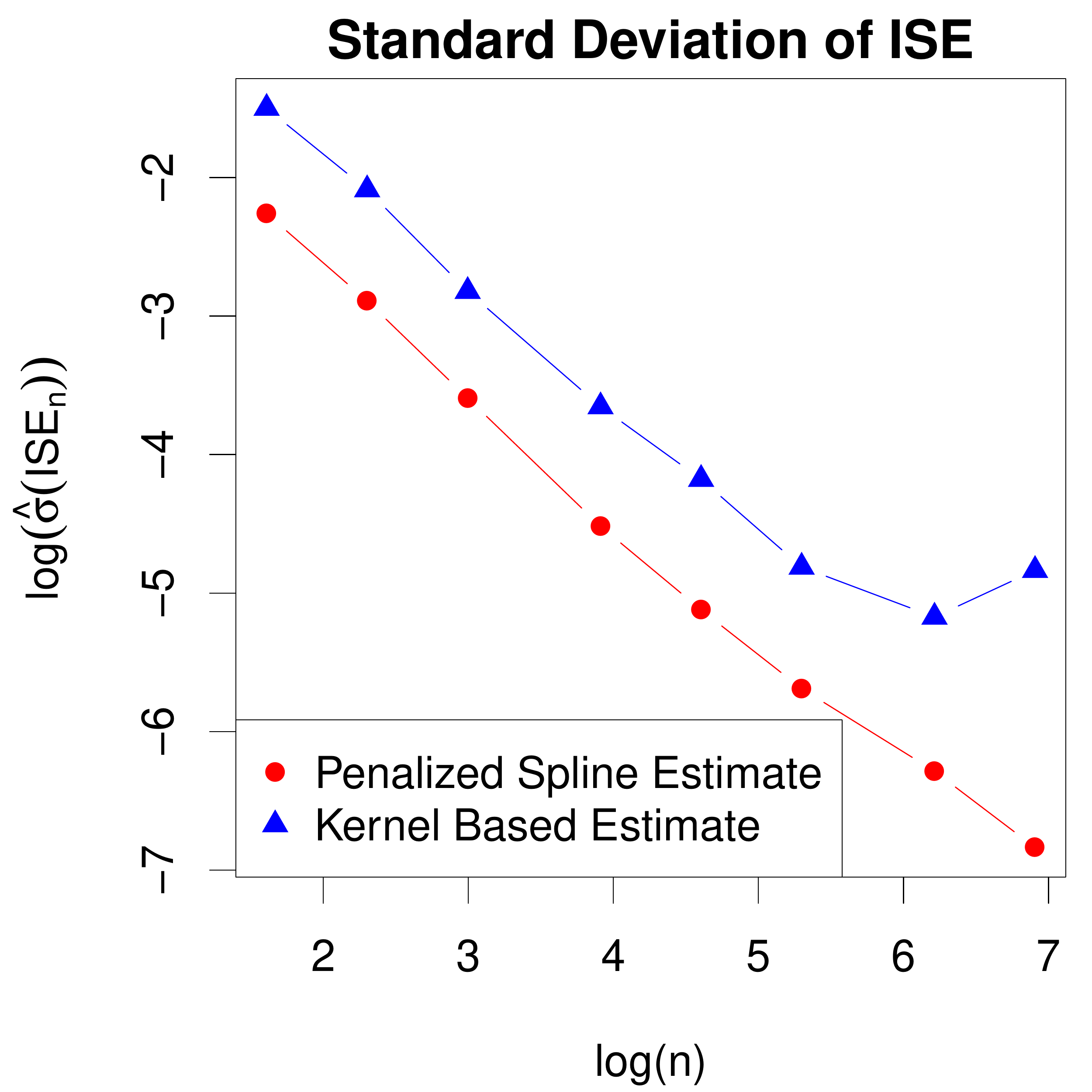}
\caption{Top panel: Plot of $\varphi_{\mathcal{X}_n}$ according to \eqref{eq: varphi exp} with $n = 1$; Mean (bottom left panel) and standard deviation (bottom right panel) of penalized spline based and kernel based $\text{ISE}$ for $n = 5,10,20,50,100,200,500,1000$ with true intensity \eqref{eq: varphi exp} on a log-log-scale.}
\label{fig: ISE exp n}
\end{figure}

\begin{table}[t]
\center
\begin{tabular}{rllll}
\toprule
 \textbf{Non-Uniform Intensity} & $h = 0.1$  & $h = 0.05$ & $h = 0.01$ & $h = 0.005$ \\ 
 \midrule
 $\delta = 0.1$  & 7.67 (5.33)  & 7.89 (5.46) & 7.73 (5.25) & 7.92 (4.93) \\
 $\delta = 0.05$ &  --          & 7.51 (4.36) & 7.62 (4.67) & 7.85 (5,40) \\
 $\delta = 0.01$ &  --          &     --      & 6.81 (4.18) & 7.78 (5.16) \\
 \bottomrule
\end{tabular}
\caption{Mean (standard deviation) of $\text{ISE}$ with $\varphi_{\mathcal{X}_n}$ as in \eqref{eq: varphi exp} for $n = 100$ as well as different choices of $\delta$ and $h$.  All values in this table are multiplied by factor 1,000.}
\label{tab: ISE exp delta h} 
\end{table}

We extend the above setting with multiple simulations per sample size and set $\delta = 0.05, h = 0.01, r = 1$. We also use several values of $n$ to assess consistency. To do so we simulate $S =  1000$ networks for $n = 5,10,20,50,100,200,500,1000$ and quantify the estimation error of the $s$-th simulation through 
\begin{align*}
\text{ISE}_n(s) = \frac{1}{n^2} \int_\mathbold{L} \left(  \varphi_{{\cal X}_n}(\mathbold{z}) - \widehat{\varphi}_{{\cal X}_n}(\mathbold{z}; s) 
\right) ^2 \dif \mathbold{z},
\label{eq: bias}
\end{align*}
where $\widehat{\varphi}_{\mathcal{X}_n}(\cdot;s)$ denotes the estimate of $\varphi_{\mathcal{X}_n}$ based on the $s$-th sample. That is, we quantify  the estimation error through the integrated squared error (ISE) between the true density $f_\mathcal{X}(\mathbold{z}) = \varphi_{\mathcal{X}_n}(\mathbold{z})/n = 1/|\mathbold{L}|$ and the estimated density $\widehat{f}_\mathcal{X}(\mathbold{z}) = \widehat{\varphi}_{\mathcal{X}_n}(\mathbold{z})/n$. The resulting means $\overline{\text{ISE}_n}$ over all $S = 1000$ simulations  and the corresponding standard deviations $\widehat{\sigma}(\text{ISE}_n)$ for penalized spline and kernel based estimation are shown in Figure \ref{fig: ISE uniform n}. We can clearly see that the penalized spline approach performs distinctly superior to the kernel based approach in terms of ISE. Even more, for $n = 1,000$ sample points we see that the latter method does perform poorly. Regarding the bottom right panel of Figure \ref{fig: simplenet}, we suppose that this results from very small bandwidths that are estimated from the data. For the penalized spline approach we see that the mean and the standard deviation of $\text{ISE}_n$ decrease nearly linearly on a log-log-scale with the number of random points $n$.

We also explore the effect of the bin width $h$ and the dimension $J$ of the B-spline basis, which is determined by the knot distance $\delta$. We only consider cases where $h \leq \delta$ and restrict this analysis on the sample size $n=100$. The results in Table \ref{tab: ISE uniform delta h} show that the mean ISE over $S = 1000$ simulations does not vary significantly with different choices of $\delta$ and $h$. This is in line with the motivating arguments in \cite{eilers1996flexible} and corresponds to the general results for penalized spline smoothing as derived in \cite{kauermann2011data}.

\subsection{Non-Uniform Intensity -- Integrated Squared Error} 

In order to investigate the performance of penalized spline based intensity estimation also for non-uniform intensities, we consider now the intensity function 
\begin{equation}
\varphi_{\mathcal{X}_n}(\mathbold{z}) = \sqrt{y} \exp(-xy)\cdot \frac{n}{C}
\label{eq: varphi exp}
\end{equation}
on the above network, which is shown in the top panel of Figure \ref{fig: ISE exp n}. Here, $\mathbold{z} = (x,y)^\top$ is the plane-coordinate representation of a point $\mathbold{z} \in \mathbold{L}$ with $0 \leq x,y \leq 1$ and $C \approx 1.558$ is the normalization constant such that $\int_\mathbold{L} \varphi_{\mathcal{X}_n} \dif \mathbold{z} = n$. In the bottom row of Figure \ref{fig: ISE exp n} we oppose the means (bottom left panel) and standard deviations (bottom right panel) of $S = 1000$ ISE samples for both, penalized spline and kernel based intensity estimation using the same parameter setting as in the uniform case from above. We can see that also here the penalized spline method is distinctly superior to the kernel based approach in terms of ISE. Furthermore, the decrease of the mean ISE is still linear on a log-log-scale but the relationship is not as strong as in the uniform case. Finally, we repeat the simulation study with $\varphi_{\mathcal{X}_n}$ as in \eqref{eq: varphi exp}, $S = 1000$ simulations, $n = 100, r = 1$ and several choices for $\delta$ and $h$. Again, there is no recognizable effect of the overall knot distance $\delta$ and the overall bin width $h$ on the mean ISE, which can be observed from Table \ref{tab: ISE exp delta h}.

\section{Application}
\label{sec: application}

\subsection{Crimes in a District of Chicago}

\begin{figure}[t]
\center
\includegraphics[width = 0.4\textwidth]{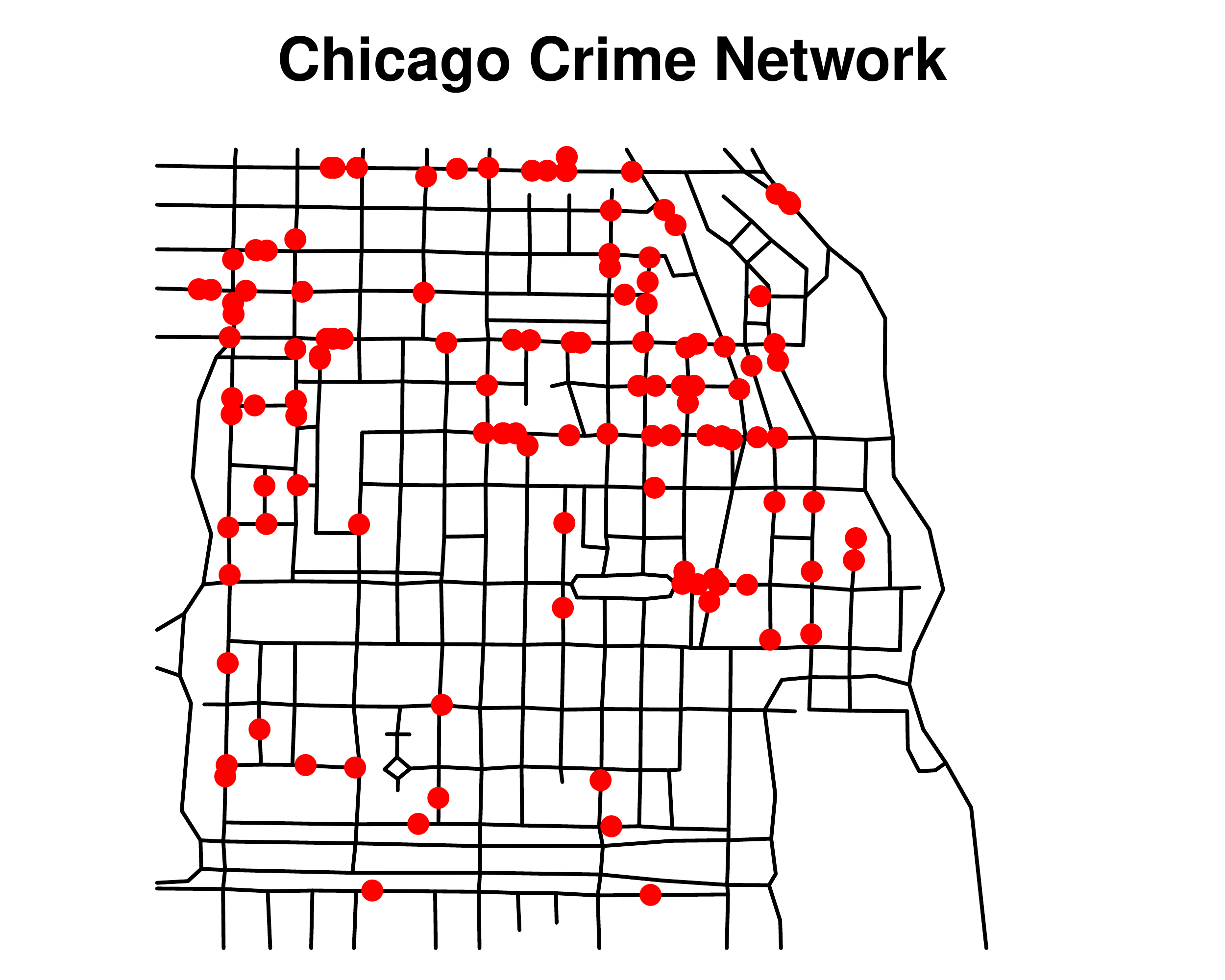} \\
\includegraphics[width = 0.4\textwidth]{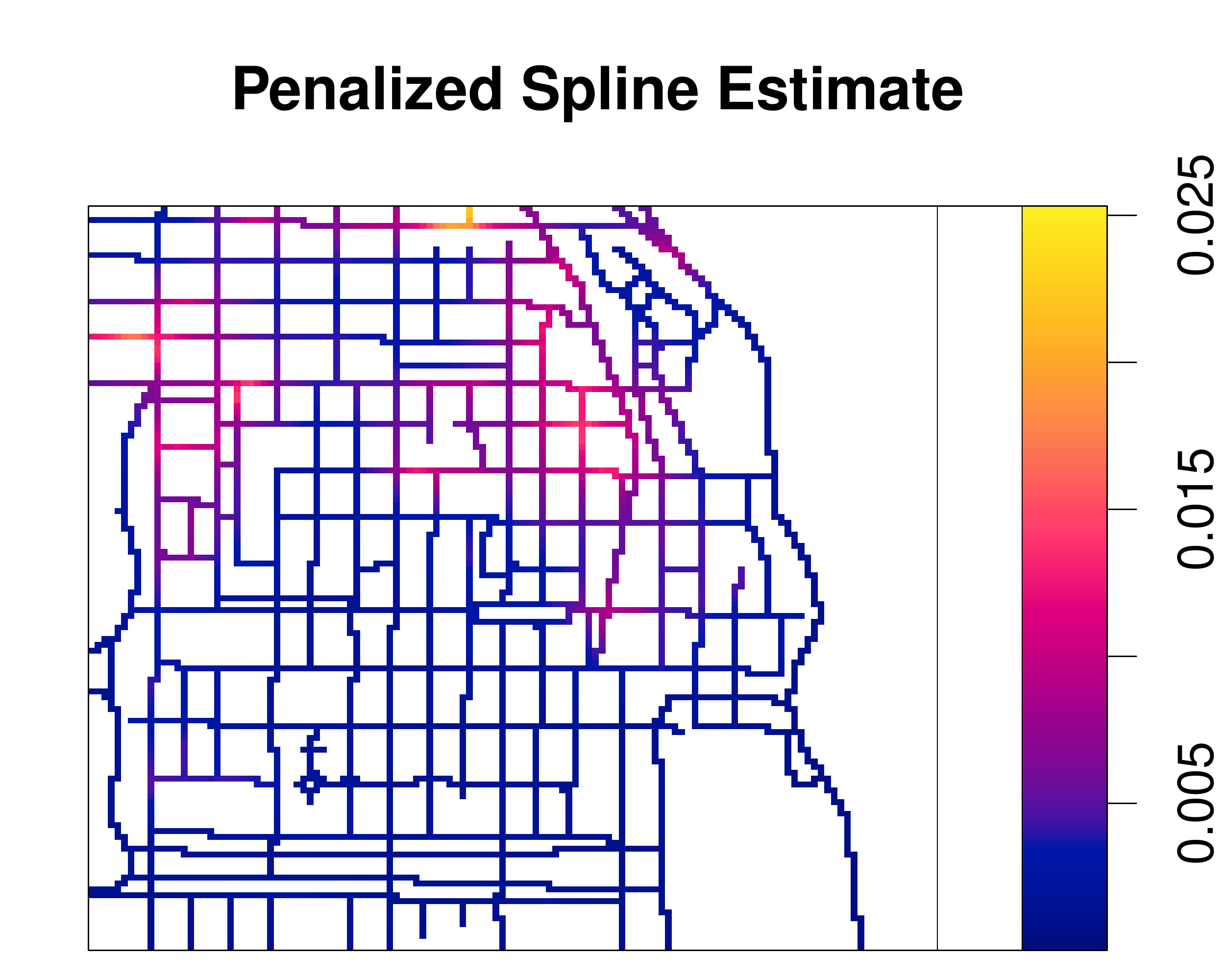}
\includegraphics[width = 0.4\textwidth]{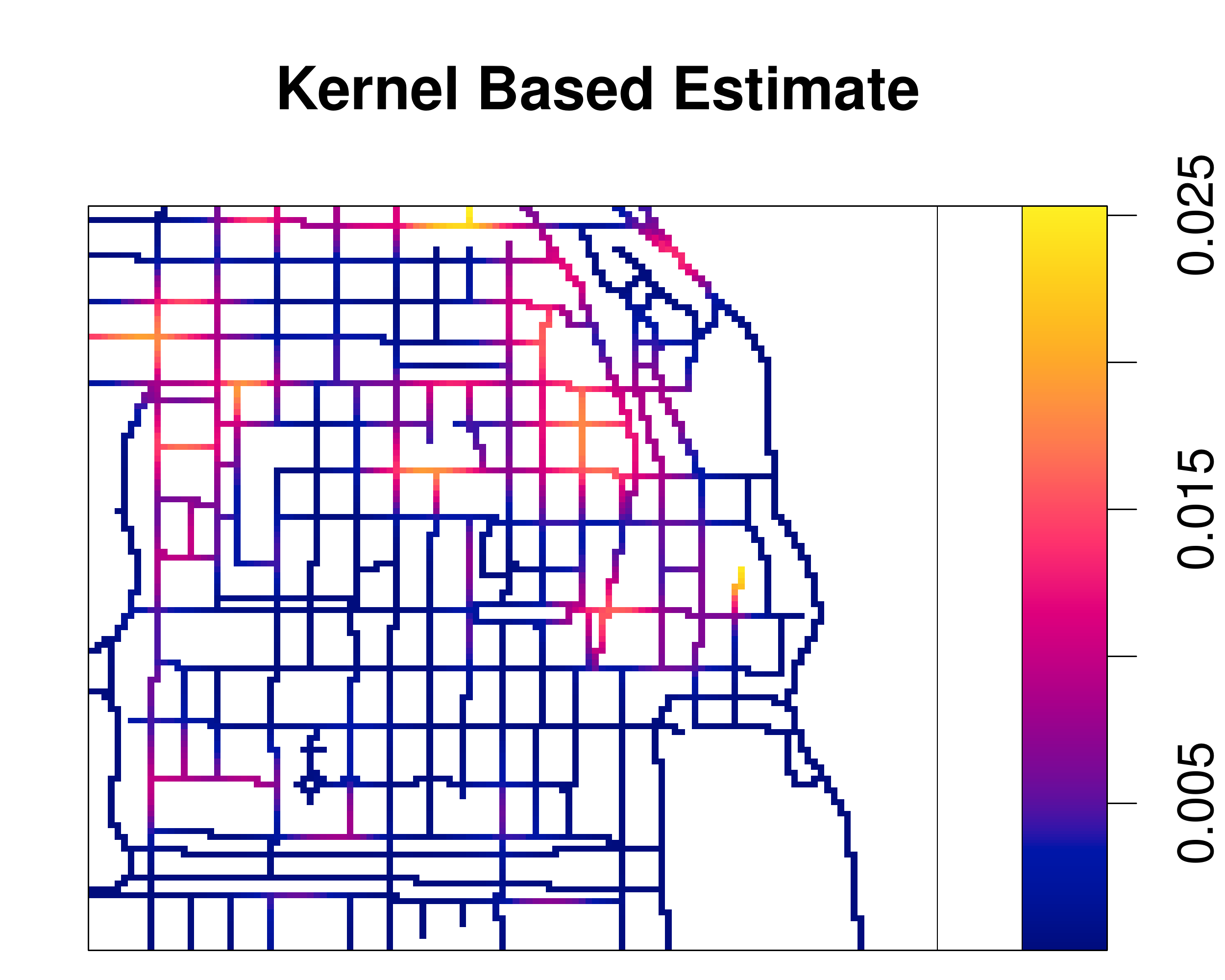}
\caption{Chicago crime network (top panel); Penalized spline intensity estimate (bottom left panel) and kernel based intensity estimate (bottom right panel).}
\label{fig: chicago}
\end{figure}

As first application with real data we consider the Chicago crimes network which is also implemented in the \texttt{spatstat} package and elaborated in detail in \cite{baddeley2015spatial}. The top panel of Figure \ref{fig: chicago} shows the location of 116 crimes recorded over a two-week period in 2002 in a district of Chicago (neglecting the kind of crime). This network consists of $|\mathbold{E}| = 503$ line segments and $|\mathbold{V}| = 338$ vertices with a total length of $|\mathbold{L}| = 31,150.21$ feet. Even though the sample size $n = 116$ is small for this rather large network, it seems very unlikely that the point pattern, which is displayed in the top panel of Figure \ref{fig: chicago}, originates from a uniform distribution on $\mathbold{L}$. Most of the crimes seem to occur in the northeastern and northwestern parts of the map extract. 

When estimating the intensity with the approach of Section \ref{sec: methodology}, we use $\delta = 5, h = 1$ and a first-order penalty. The lower panels of Figure \ref{fig: chicago} show the estimates resulting from the data using a penalized spline based approach (left panel) or a kernel based estimation (right panel). We find  that the high-intensity regions are similarly located for both methods. However, the peaks of the kernel intensity estimate are more explicit and the intensity fitted with the penalized spline approach is more smooth. When considering the few events in the southern area of the map extract, a peak occurs in the kernel intensity estimate which is not visible in the spline approach. Taking the performance of the kernel density estimates as shown in the simulated data above into account, it seems plausible to consider crime events to be uniformly distributed over the southern part of the network, that is, the peaky behavior of the kernel methods seems misleading.

\subsection{On-Street Parking in Melbourne}

Between August 2011 and May 2012, the City of Melbourne, Australia, installed in-ground sensors underneath around $4,600$ out of more than $20,000$ on-street parking lots in the Central Business District (CBD) of Melbourne. These sensors are able to record the arrival time and the departing time of a car to the second.\footnote{The massive amounts of data that these sensors produce and many more data related to parking in the City Melbourne are available for gratuitous download at \url{https://data.melbourne.vic.gov.au/}.} Our goal is to detect regions in the CBD where the occupation of on-street parking lots fluctuates most. Therefore, we define an event to be the clearing of an on-street parking lot and the point process that we observe now has a spatial as well as a temporal structure. We here concentrate on the spatial structure only.

\begin{figure}[t]
\center
\includegraphics[width = 0.4\textwidth]{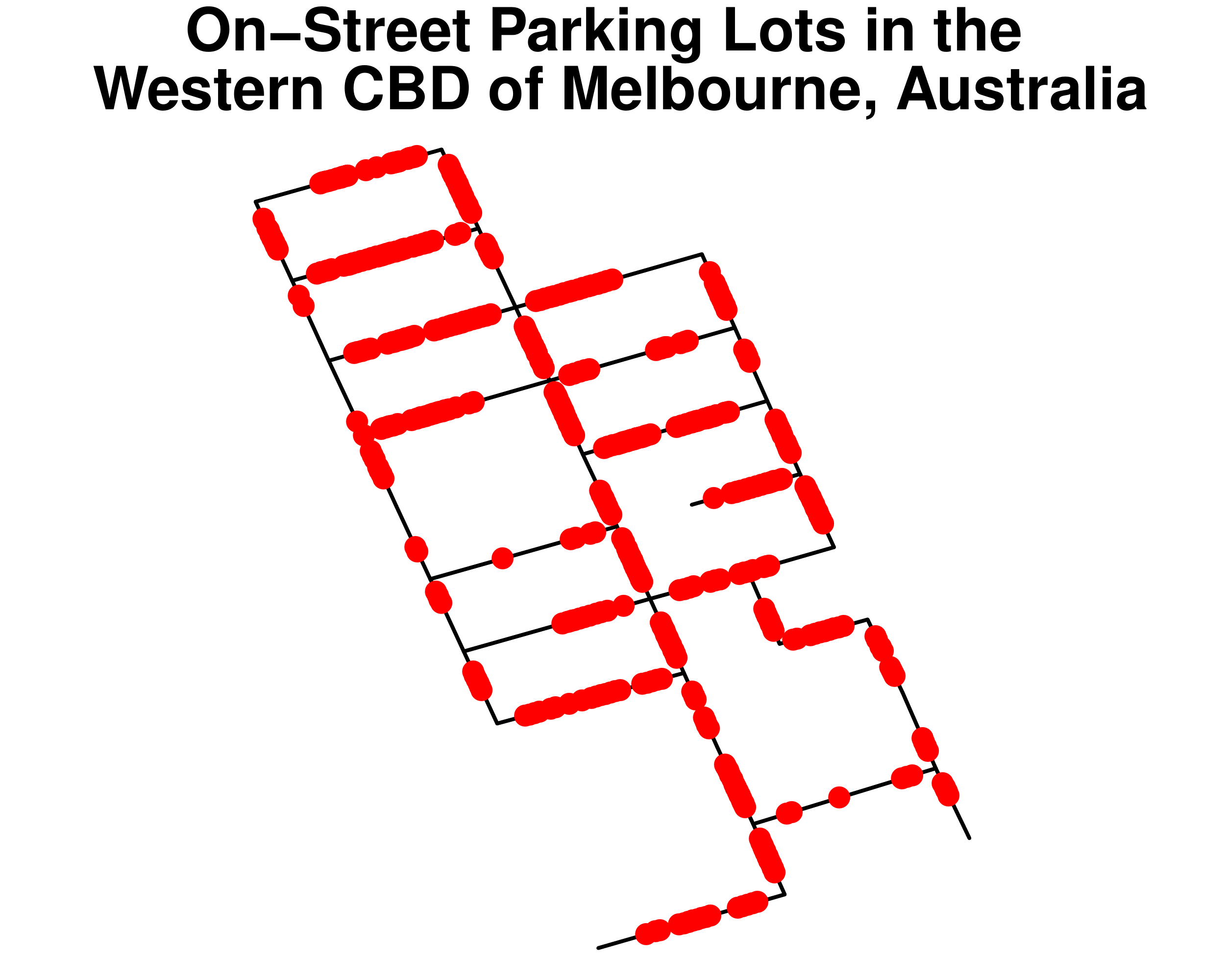} \hspace{1cm}
\includegraphics[width = 0.4\textwidth]{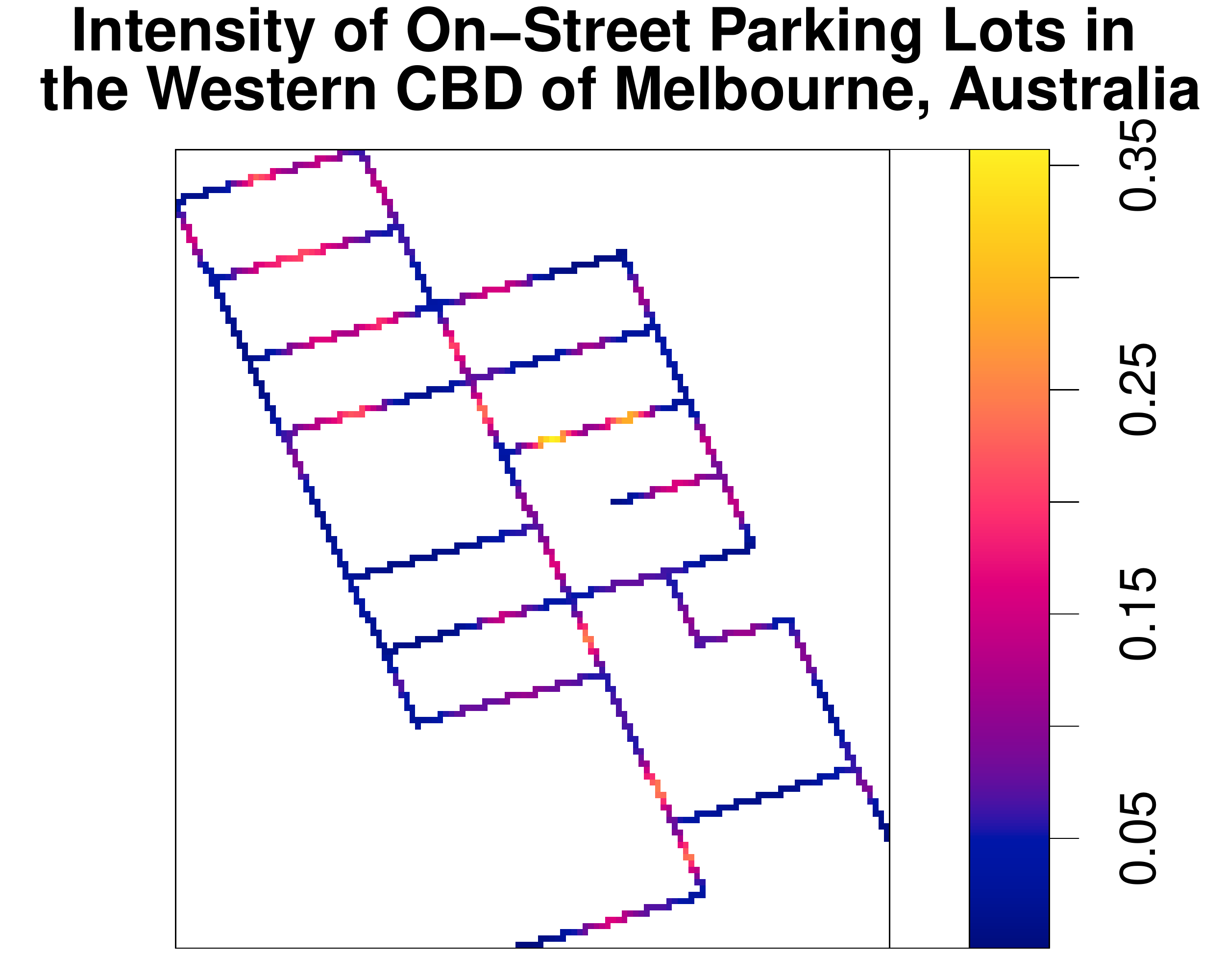}
\caption{The western CBD of Melbourne, Australia: Parking bays with installed in-ground sensors within the specified geometric network (left panel); Penalized spline estimate of the intensity of parking lots on this network (right panel).}
\label{fig: melbourne lots}
\end{figure}

For reasons of simplicity, we only consider a subset of 518 on-street parking lots with installed in-ground sensors, which are all located in the western CBD of Melbourne. To estimate the fluctuation rate, we first need to specify a geometric network where the point process is living on. This network is constructed by excluding edges (i.e. streets in this area) without installed parking lots, since by definition we can not observe parking events on these streets. In Figure \ref{fig: melbourne lots} we show the location of the considered on-street parking lots on the corresponding geometric network, which consists of $|\mathbold{E}| = 41$ line segments, $|\mathbold{V}| = 32$ vertices and has a length of nearly 7 kilometers. In order to estimate the fluctuation rate, we also need to estimate the intensity $\varphi_\mathcal{Z}$ of parking lots along the geometric network $\mathbold{L}$. The fitted intensity $\widehat{\varphi}_{\mathcal{Z}}$ of allocated parking lots using the penalized spline based approach is depicted in the right panel of Figure \ref{fig: melbourne lots}. Note that this estimate is not based on data on cleared parking lots but on the existence of the lots itself. In other words, this estimate may serve as baseline subsequently, since in areas with more allocated parking lots there is per se a higher chance of finding a cleared lot.

\begin{figure}[t]
\center
\includegraphics[width = 0.4\textwidth]{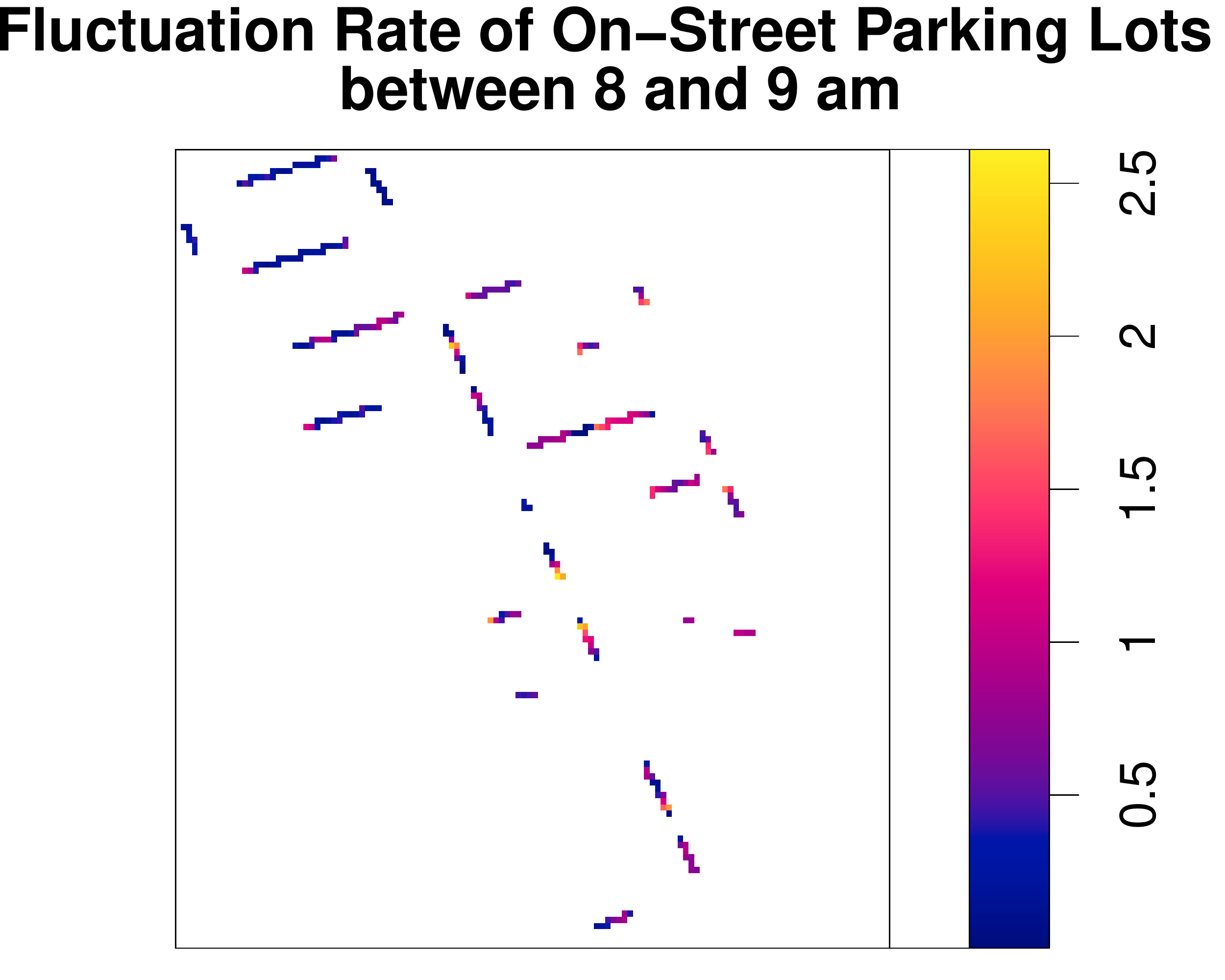} \hspace{1cm}
\includegraphics[width = 0.4\textwidth]{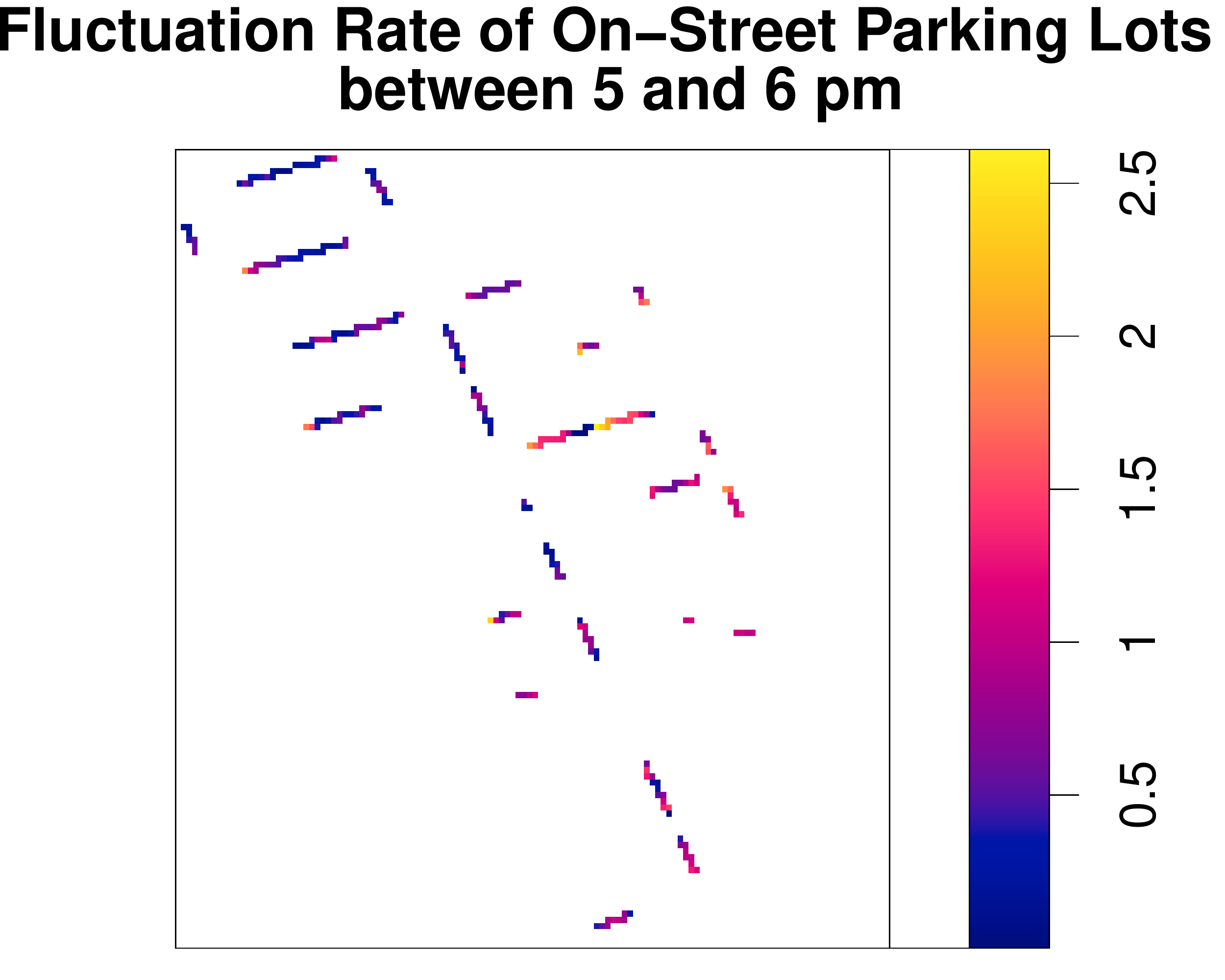}
\caption{Fluctuation rate (per hour) of on-street parking lots in the western CBD of Melbourne where we expect at least one parking lot per 10 meters: Between 8 and 9 am (left panel); Between 5 and 6 pm (right panel).}
\label{fig: melbourne intensity}
\end{figure}

We now look at cleared parking lots and consider data in the hourly intervals between 8 and 9 am in the morning as well as 5 and 6 pm in the evening. We take data on working days over the year 2017 and consider these data as results of the clearing point process $\mathcal{Y}$, whose intensity $\varphi_\mathcal{Y}$ is again estimated using penalized spline smoothing. Overall we are interested in the ratio $\varphi_\mathcal{X} = \varphi_\mathcal{Y}/\varphi_\mathcal{Z}$, which expresses the expected fluctuation rate of parking lots along the network. However, we are only interested in the fluctuation rate where we expect a reasonable number of parking lots. Therefore, we show in Figure \ref{fig: melbourne intensity} the estimated fluctuation rates $\widehat{\varphi}_\mathcal{X}(\mathbold{z})$ only at locations $\mathbold{z} \in \mathbold{L}$ where $\widehat{\varphi}_\mathcal{Z}(\mathbold{z}) \geq 0.1$, i.e. where we expect at least one on-street parking lot per 10 meters. Furthermore, the rates are normalized such that $\widehat{\varphi}_\mathcal{X}(\mathbold{z})$ can be interpreted as the expected hourly fluctuation rate of a parking lot, which is located at $\mathbold{z} \in \mathbold{L}$.

When we examine Figure \ref{fig: melbourne intensity}, we can see that the high intensity regions distinguish between the hour in the morning (left panel) and the hour in the evening (right panel). In the former case, the highest fluctuation rate with around 3 clearances per parking lot and hour can be observed in parts of the central north/south road (King Street) of this map extract. In the latter case, we see a much lower fluctuation rate in King Street but a much higher fluctuation rate in the northwestern part of the map extract, especially in Lonsdale Street between King Street and William Street. Here, we have the highest intensity of parking lots $\varphi_\mathcal{Z}$ but also numerous clearing events that we observe between 5 and 6 pm.

\section{Discussion and further Work}
\label{sec: discussion}

In this article we developed a new method for the estimation of the intensity (or density) of a stochastic process living on a geometric network. We exploited and extended penalized spline estimation to work on a subset of connected curves which we denoted as geometric networks. Note that by our definition, an interval $[a,b]$ is a special case of a geometric network $\mathbold{L}$ with $|\mathbold{E}| = 1$ and $|\mathbold{V}| = 2$. Our results show that penalized spline based smoothing on geometric networks benefits over the current state of the art estimation based on kernel smoothing.

As seen in the simulation study and the application examples, the penalization also compensates for non-equidistant knots and bin widths on different segments of $\mathbold{L}$. However, these differences can be made as small as desired by reducing $\delta$ and $h$. In the end, this leads to a trade-off between accuracy and computational effort. In an Euclidean space, the penalties that are used for estimation with B-splines are often based on derivatives of the smoother. However, in a geometric network the question arises how one could define differentiability of a function $f$ at vertices $\mathbold{v}$ with $\text{deg}(\mathbold{v}) > 2$. Apparently, adapting the penalization technique of \cite{eilers1996flexible} circumvents this question and proves to be the right choice for our setting.

We envisage many more generalizations and extensions of our method. First the linear penalized spline approach could be extended to work with higher order penalized splines, in particular with quadratic or cubic penalized splines. Therewith, the estimated intensities could become even more smoother along the network. However, B-splines of order 2 or higher in Euclidean spaces are differentiable. Therefore, as stated above it would be much more complicated to construct network based B-splines of order 2 or higher around vertices $\mathbold{v}$ with $\text{deg}(\mathbold{v}) > 2$.

Furthermore, if we drop the assumption that the network graph $L$ should not be directed, we need the geometric representation $\mathbold{L}$ to be possibly directed as well. This means that a curve $\mathbold{e}_m$ additionally is equipped with a direction if $e_m = (v_i, v_j)$ is a directed edge from $v_i$ to $v_j$ but there is no edge from $v_j$ to $v_i$. In this case, the distance measure $d_\mathbold{L}$ from above does not define a metric any more since then, $d_\mathbold{L}(\mathbold{z_1}, \mathbold{z}_2) = d_\mathbold{L}(\mathbold{z_2}, \mathbold{z}_1)$ for $\mathbold{z_1}, \mathbold{z_2} \in \mathbold{L}$ does not hold in general.  

Apart from estimating intensities, our construction of linear B-splines on a geometric network also allows to include a network specific covariate in a regression model. For example, the time until a parking lot is reoccupied after a car has left can be explained by covariates such as time of the day, weather conditions, etc. But also the location of the parking lot in the street network might be an explanatory variable which can be smoothly modeled with our penalized spline approach. 

And finally, the time aspect can be incorporated as well, leading to simultaneous smoothing over the geometric network and time. We consider these aspects beyond the scope of the current paper but see high potential for future research. For example, it would be interesting how the fluctuation rates in the Melbourne parking network vary during the course of the year.

\FloatBarrier

\bibliographystyle{Chicago}
\bibliography{literature}

\end{document}